\documentclass{aa}  

\usepackage{graphicx}
\usepackage{txfonts}
\usepackage{verbatim} 
\usepackage[normalem]{ulem}

\usepackage[dvipsnames]{xcolor}
\usepackage{soul}
\usepackage{bm}

\def\co2{CO$_2$}
\def\ch4{CH$_4$}
\def\h2{H$_{2}$}
\def\h2o{H$_2$O}
\def\nh3{NH$_3$}
\def\hdu18{HD~189733~b}
\def\hd20{HD~209458~b}

\def\kms{km\,s$^{-1}$}
\def\kp{$K_{\rm P}$}

\begin{document} 

\titlerunning{Water vapor detection in the transmission spectra of \hd20 with the CARMENES NIR channel}
\authorrunning{A. S\'anchez-L\'opez, et al.}
\title{Water vapor detection in the transmission spectra of \hd20 with the CARMENES NIR channel}
%
%
%
\author{A. S\'anchez-L\'opez\inst{1}, F. J. Alonso-Floriano\inst{2}, M. L\'opez-Puertas\inst{1}, I. A. G. Snellen\inst{2}, B. Funke\inst{1}, E. Nagel\inst{3},  F.~F.~ Bauer\inst{1},  P.~J. Amado\inst{1}, J.~A. Caballero\inst{4}, S. Czesla\inst{3}, L. Nortmann\inst{5,6}, E. Pall{\'e}\inst{5,6}, M. Salz\inst{3}, A. Reiners\inst{7}, I. Ribas\inst{8,9},  A.~Quirrenbach\inst{10}, G. Anglada-Escud{\'e}\inst{11, 1}, V.~J.~S.~B{\'e}jar\inst{5,6}, N. Casasayas-Barris\inst{5, 6}, D. Galad\'{\i}-Enr\'{\i}quez\inst{12}, E.~W.~Guenther\inst{13}, Th. Henning\inst{14},   A.~Kaminski\inst{10}, M. K{\"u}rster\inst{14}, M.~Lamp{\'o}n\inst{1}, L. M. Lara\inst{1}, D.~Montes\inst{15}, J.~C.~Morales\inst{8,9}, M.~Stangret\inst{5, 6}, L. Tal-Or\inst{16, 7}, J. Sanz-Forcada\inst{4}, J. H. M. M. Schmitt\inst{3}, M.~R.~Zapatero~Osorio\inst{17}, and M. Zechmeister\inst{7}}
\institute{Instituto de Astrof{\'i}sica de Andaluc{\'i}a (IAA-CSIC), Glorieta de la Astronom{\'i}a s/n, 18008 Granada, Spain\\
\email{alexsl@iaa.es}
\and
Leiden Observatory, Leiden University, Postbus 9513, 2300 RA, Leiden, The Netherlands
\and
Hamburger Sternwarte, Universit{\"a}t Hamburg, Gojenbergsweg 112, 21029 Hamburg, Germany
\and
Centro de Astrobiolog{\'i}a (CSIC-INTA), ESAC, Camino bajo del castillo s/n, 28692 Villanueva de la Ca{\~n}ada, Madrid, Spain
\and
Instituto de Astrof{\'i}sica de Canarias (IAC), Calle V{\'i}a L{\'a}ctea s/n, 38200 La Laguna, Tenerife, Spain
\and
Departamento de Astrof{\'i}sica, Universidad de La Laguna, 38026  La Laguna, Tenerife, Spain
\and
Institut f{\"u}r Astrophysik, Georg-August-Universit{\"a}t, Friedrich-Hund-Platz 1, 37077 G{\"o}ttingen, Germany
\and
Institut de Ci\`encies de l'Espai (CSIC-IEEC), Campus UAB, c/ de Can Magrans s/n, 08193 Bellaterra, Barcelona, Spain
\and
Institut d'Estudis Espacials de Catalunya (IEEC), 08034 Barcelona, Spain
\and
Landessternwarte, Zentrum f\"ur Astronomie der Universit\"at Heidelberg, K\"onigstuhl 12, 69117 Heidelberg, Germany
\and
School of Physics and Astronomy, Queen Mary, University of London, 327 Mile End Road,
London, E1 4NS, UK
\and
Observatorio de Calar Alto, Sierra de los Filabres, 04550 G\'ergal, Almer\'{\i}a, Spain
\and
Th{\"u}ringer Landessternwarte Tautenburg, Sternwarte 5, 07778 Tautenburg, Germany
\and
Max-Planck-Institut f{\"u}r Astronomie, K{\"o}nigstuhl 17, 69117 Heidelberg, Germany
\and
Departamento de F{\'i}sica de la Tierra y Astrof{\'i}sica \& IPARCOS-UCM (Instituto de F{\'i}sica de Part{\'i}culas y del Cosmos de la UCM), Facultad de Ciencias F{\'i}sicas, Universidad Complutense de Madrid,  28040 Madrid, Spain
\and
Department of Geophysics, Raymond and Beverly Sackler Faculty of Exact Sciences, Tel Aviv University, Tel Aviv 6997801, Israel
\and
Centro de Astrobiolog{\'i}a (CSIC-INTA), Carretera de Ajalvir km 4,
28850 Torrej{\'o}n de Ardoz, Madrid, Spain
}

\date{}

 
\abstract
{}
{We aim at detecting water vapor in the atmosphere of the hot Jupiter \hd20 and perform a multi-band study in the near infrared with CARMENES.}
{The water vapor absorption lines from the atmosphere of the planet are Doppler-shifted due to the large change in its radial velocity during transit. This shift is of the order of tens of km\,s$^{-1}$, whilst the Earth's telluric and the stellar lines can be considered quasi-static. We took advantage of this shift to remove the telluric and stellar lines using {\sc Sysrem}, which performs a principal component analysis including proper error propagation. The residual spectra contain the signal from thousands of planetary molecular lines well below the noise level. We retrieve the information from those lines by cross-correlating the residual spectra with models of the atmospheric absorption of the planet.}
{We find a cross-correlation signal with a signal-to-noise ratio (S/N) of 6.4, revealing \h2o in \hd20. We obtain a net blueshift of the signal of --5.2\,$^{+2.6}_{-1.3}$\,km\,s$^{-1}$ that, despite the large error bars, is a firm indication of day- to night-side winds at the terminator of this hot Jupiter. Additionally, we performed a multi-band study for the detection of \h2o individually from the three near infrared bands covered by CARMENES. We detect \h2o from its 0.96--1.06\,$\mu$m band with a S/N of 5.8, and also find hints of a detection from the 1.06--1.26\,$\mu$m band, with a low S/N of 2.8. No clear planetary signal is found from the 1.26--1.62\,$\mu$m band. 
}
{Our significant \h2o signal at 0.96--1.06\,$\mu$m in \hd20 represents the first detection of \h2o from this band individually, the bluest one to date. The unfavorable observational conditions might be the reason for the inconclusive detection from the stronger 1.15 and 1.4\,$\mu$m bands. \h2o is detected from the 0.96--1.06\,$\mu$m band in \hd20, but hardly in \hdu18, which supports a stronger aerosol extinction in the latter, in line with previous studies. Future data gathered at more stable conditions and with larger S/N at both optical and near-infrared wavelengths could help to characterize the presence of aerosols in \hd20 and other planets.}


   \maketitle
%

\section{Introduction}
\label{introduction}

Transmission spectroscopy during the primary transit of exoplanets has proven to be extremely useful for characterizing their atmospheres \citep{Seager00, Brown01, Hubbard01}. As the planet passes in front of its host star from the perspective of the Earth, part of the stellar light is blocked by the planetary disk, opaque at all wavelengths. In addition, part of the light is absorbed and scattered by the constituents of the planet's atmosphere, which is seen as an annulus from the point of view of the observer. Each compound absorbs in specific spectral regions and with different intensities, which leaves weak but unique fingerprints in the transmitted light. Taking advantage of this, the first detection of an exoplanet atmosphere was achieved by \cite{Charbonneau02}, who reported a detection of sodium in the atmosphere of \hd20 using the Space Telescope Imaging Spectrograph on the \textit{Hubble Space Telescope} ($HST$/STIS). Only a year later, \cite{VidalMadjar2003} detected the absorption of hydrogen in Lyman-$\alpha$ in the atmosphere of the same exoplanet, also using $HST$/STIS. Later, \cite{Redfield08} and \cite{Snellen08} accomplished the first detections of planetary sodium with ground-based instrumentation at 8--10\,m class telescopes. 

The signatures produced by molecular species in the planet's atmosphere are also orders of magnitude weaker than those of the telluric and stellar lines. This greatly hinders their detection from the ground since their spectral features cannot be seen directly in the spectra, but are buried well below the noise level. To overcome this difficulty, \cite{Snellen10} merged the information of thousands of CO rotational-vibrational (ro-vibrational) lines in \hd20 transit spectra by applying a cross-correlation technique to high-resolution spectroscopy with the Cryogenic InfraRed \'Echelle Spectrograph (CRIRES) at the Very Large Telescope. \cite{Snellen10} took advantage of the Doppler shift between the stellar light and the planetary atmospheric absorption caused by the planet's movement around its host star. This is possible because the semi-amplitude of the planet's orbital velocity can be well beyond 100\,km\,s$^{-1}$ and the relative velocity with respect to the Earth changes by tens of km\,s$^{-1}$ during the transit. On the contrary, the Earth's telluric lines and the stellar lines can be considered quasi-static. Moreover, \cite{Snellen10} measured an excess Doppler-shift of 2\,$\pm$\,1\,km\,s$^{-1}$, which was explained by invoking high-altitude winds in the atmosphere. Moreover, it constituted the first evidence of high-resolution instrumentation being sensitive to the atmospheric dynamics.

Ever since, the field of atmospheric characterization using cross-correlation and high-resolution spectroscopy ($\mathcal{R}\sim$\,100\,000) with ground-based instruments at 8--10\,m class telescopes has flourished. For instance, the signatures of \h2o, CO, Fe, and TiO have been identified in several transiting \citep{Birkby13, deKok13, Schwarz15, Brogi16, Nugroho17, Hoeijmakers18} and non-transiting planets \citep{Brogi12, Brogi13, Brogi14, Lockwood14, Piskorz16, Piskorz17, Birkby17}. 

The cross-correlation technique has also been successfully applied to mid- and high-resolution ($\mathcal{R}\geq$\,50\,000) highly-stabilized instruments at 4--m class telescopes for detecting molecular signatures. Thus, water vapor was detected in HD\,189733\,b by \cite{Brogi18} using GIANO at the Telescopio Nazionale Galileo. They reported a blueshift in the signal of $-$1.6\,km\,s$^{-1}$, indicating the presence of winds in the atmosphere. 
Moreover, \cite{Alonso19} detected \h2o in the same exoplanet for the first time using the 1.15 and 1.4\,$\mu$m bands individually with the near-infrared (NIR) channel of CARMENES \cite[Calar Alto high-Resolution search for M dwarfs with Exoearths with Near-infrared and optical Échelle Spectrographs;][]{Quirrenbach16, Quirrenbach18}.

One of the best-studied planetary atmospheres to date is that of \hd20, which orbits a bright Sun-like star (G0\,V, $G$\,$\approx$\,7.5\,mag) with a period of 3.5 days \citep{Charbonneau00, Henry00}. This planet has been studied by analyzing its transit \citep{Charbonneau02, VidalMadjar2003, Deming13, Madhusudhan2014a} and dayside spectra \citep{Line16, Hawker2018}, combining low- and high-resolution spectra \citep{Brogi17}, and by investigating its exosphere and mass-loss rate \citep{VidalMadjar2004, Linsky10}.

Comparative studies of several hot Jupiters using $HST$ and the \textit{Spitzer Space Telescope} have shown water vapor spectral features in \hd20 in the 0.3--5.0\,$\mu$m spectral region \citep{Sing16}. These signatures were found to be weaker than expected in a solar composition cloud-free atmosphere, and have been attributed to a possible oxygen depletion or a partial coverage of clouds or hazes \citep{Madhusudhan2014b}. Later, \cite{MacDonald17} found sub-solar abundances in the atmosphere of \hd20, with inhomogeneous clouds covering the terminator. In addition, they reported the first detection of nitrogen chemistry, with NH$_3$ (or HCN) being detected in $HST$ transmission spectra in the 1.1--1.7\,$\mu$m region. Moreover, \cite{Hawker2018} also detected HCN, \h2o, and CO in this planet by using CRIRES measurements in the 2.29--2.35\,$\mu$m and 3.18--3.27\,$\mu$m spectral ranges.

In this paper we analyze transit spectra of \hd20  obtained with the near infrared (NIR) channel of CARMENES (0.96--1.71\,$\mu$m). After the multi-band detection of water vapor in \hdu18 with this instrument \citep{Alonso19}, we have extended this analysis to \hd20, where \h2o has not been detected from the ground so far in the NIR. Detections of water vapor from bands located at different spectral regions can help to characterize aerosols in hot Jupiters \citep{Stevenson2016, Pino18b}. This kind of multi-band studies is of particular interest for \hdu18 and \hd20 since space observations suggest that the former is rather hazy whilst the latter has a more gentle Rayleigh scattering slope \citep{Sing16}. Moreover, an additional measurement of the winds in the upper atmosphere of \hd20 is also of high interest for characterizing the dynamics of hot Jupiter atmospheres. 

This paper is organized as follows. In Sect.~\ref{observations} we describe the main features of the CARMENES instrument, our observations of \hd20 and the atmospheric conditions during the night. In Sect.~\ref{datareduction} we present the data reduction procedure and in Sect.~\ref{signalretrieval} the methodology for identifying the molecular signatures. In Sect. \ref{results_discussion} we elaborate on the main results of this work regarding the \h2o detection in the different bands, the atmospheric winds, and a comparison with the results for \hdu18. Finally, in Sect. \ref{conclusions} we present the main conclusions.

\section{Observations} 
\label{observations}

We observed a transit of the hot Jupiter HD\,209458\,b (Table \ref{table.parameters}) on the night of 5 September 2018 using CARMENES. CARMENES is located in the coud\'e room of the 3.5\,m telescope at the Calar Alto Observatory. The instrument consists of two fiber-fed high-resolution spectrographs, the visible (VIS) channel covering optical wavelengths, $\Delta\lambda$\,=\,520--960\,nm in 55 orders ($\mathcal{R}$\,=\,94\,600), and the NIR channel observing the near-infrared wavelength range $\Delta\lambda$\,=\,960--1710\,nm in 28 orders ($\mathcal{R}$\,=\,80\,400). In this work we analyze the data gathered with the NIR channel of CARMENES only.

\begin{figure}
\includegraphics[angle=0, width=1.0\columnwidth]{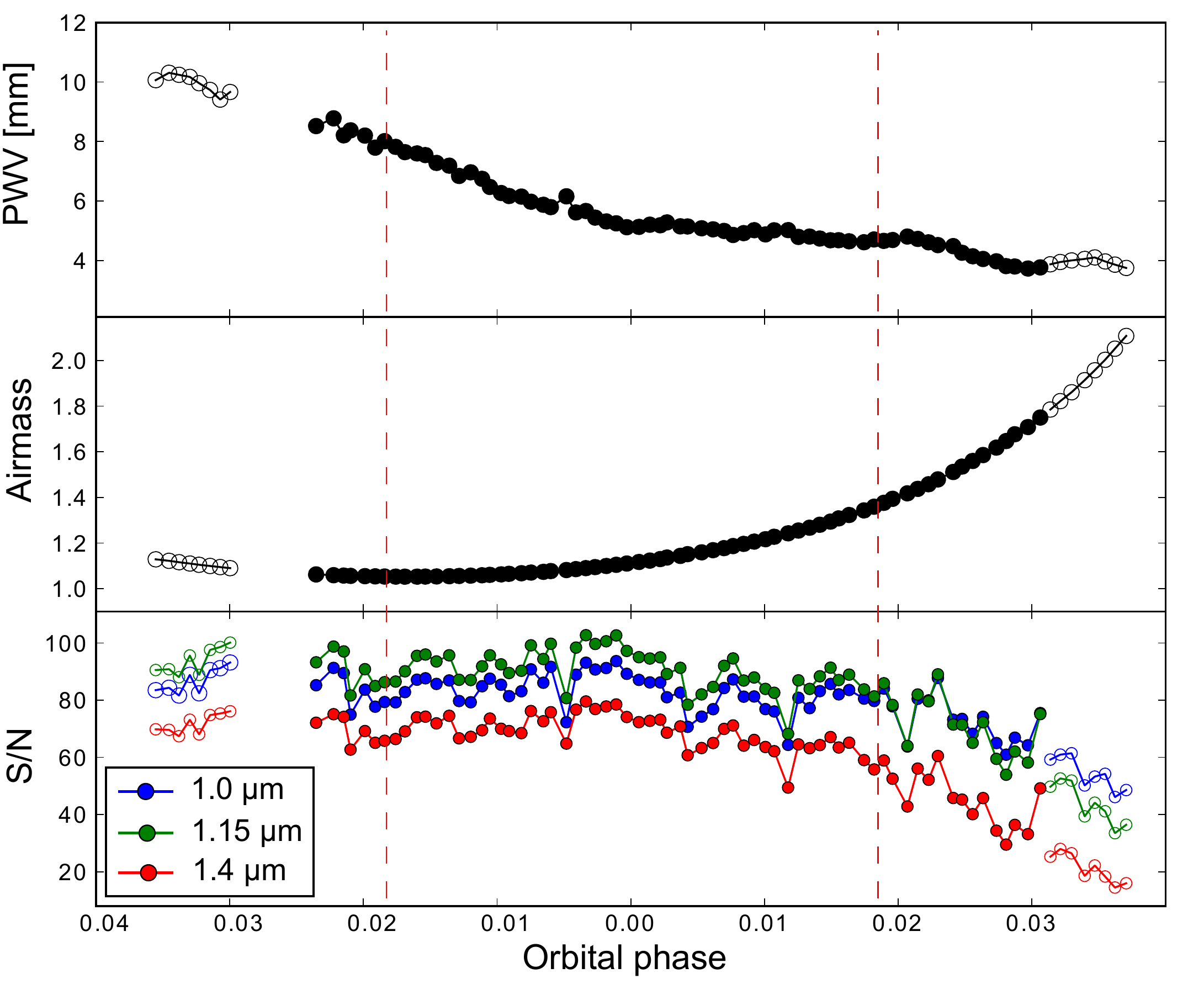}\hspace*{0.35cm}
\caption{Column depth of precipitable water vapor in the Earth's atmosphere (\textit{top panel}), airmass (\textit{middle panel}), and mean S/N of the raw spectra for the three covered \h2o bands (\textit{bottom panel}) as a function of the orbital phase. Open circles represent spectra not included in the analysis (see text). The transit occurred at the orbital phases between the vertical dashed lines.} 
\label{pwv_airmass_sn}
\end{figure}

\begin{figure}
\includegraphics[angle=0, width=1.0\columnwidth]{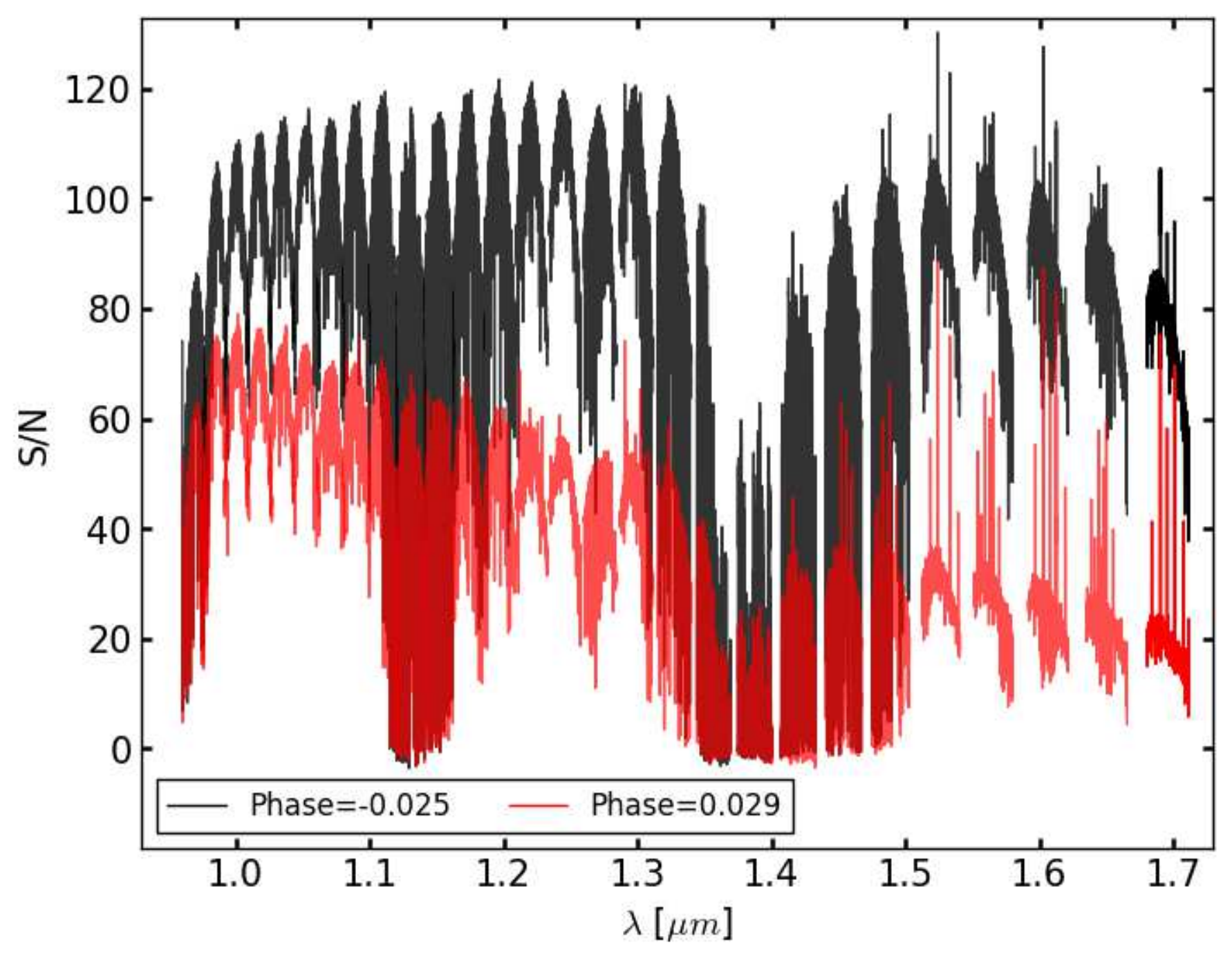}\hspace*{0.35cm}
\caption{S/N of spectra taken at the beginning (black) and near the end (red) of the observations. A strong spectral dependence of the S/N is observed in the latter.} 
\label{sn_spectra}
\end{figure}

We used the two fibers of the instrument. Fiber A was devoted to the target, and fiber B to the sky in order to identify sky emission lines.
The observations covered the pre-, in-, and post-transit of \hd20, and are composed of 91 exposures of 198\,s from 21:39 UT to 03:47 UT (see Fig.\,\ref{pwv_airmass_sn}).
Due to a brief failure in the guiding system, eight spectra were missed between the planetary orbital phases $\phi=-$0.029 and $\phi=-$0.024. We thus had to discard also the eight pre-transit spectra gathered before them in order to avoid sharp discontinuities in the light curve of each pixel, which would hamper our telluric corrections (see Sect.~\ref{tellurics}).
The typical signal-to-noise ratio (S/N) per pixel of the raw spectra showed a strong spectral dependence (see bottom panel in Fig.\,\ref{pwv_airmass_sn} and Fig.\,\ref{sn_spectra}). The mean S/N was of $\sim$\,85 for the bands at $\sim$\,1\,$\mu$m and $\sim$\,1.15\,$\mu$m and of $\sim$\,65 for the band at $\sim$\,1.4\,$\mu$m in the first half of the observations. However, it dropped below 60 towards the end for the bands at $\sim$\,1\,$\mu$m and $\sim$\,1.15\,$\mu$m, and below 50 for the band at $\sim$\,1.4\,$\mu$m. Consecutively, we also discarded the last eight out-of-transit spectra to avoid noisy data. Additionally, one in-transit spectrum presented an anomalous spectral behavior and was also excluded from the analysis.
The spectra analyzed in this work thus spanned the exoplanet orbital phase interval $-$0.023\,<\,$\phi$\,<\,0.031. The airmass range was 1.06\,<\,sec\,$\left(z\right)$\,<\,1.79 (see Fig.\,\ref{pwv_airmass_sn}) and the mean seeing was 1.5\,arcsec.

\begin{table}
\centering
\caption{\label{table.parameters}Parameters of the HD\,209458 system.}
\begin{tabular}{l c l} 
   \hline
   \hline
   \noalign{\smallskip}
Parameter					 & Value & Reference  \\
  \noalign{\smallskip}
    \hline
      \noalign{\smallskip}
$\alpha$ [J2000]	& 22:03:10.77 	& {\em Gaia} DR2$^{a}$	\\
\noalign{\smallskip}
$\delta$ [J2000]	& +18:53:03.5	& {\em Gaia} DR2$^{a}$ \\
\noalign{\smallskip}
$G$		& $7.5087\,(4)$\,mag 	&	{\em Gaia} DR2$^{a}$ \\
\noalign{\smallskip}
$J$			& $6.59\,(2)$\,mag 	& \cite{Skrutskie06}	 \\
\noalign{\smallskip}
$R_{\star}$$^{b}$ &	1.155\,$^{+0.014}_{-0.016}$\,R$_\sun$	 & \citet{Torres08}	\\
\noalign{\smallskip}
$K_{\star}$ & $84.67(70)$\,m\,s$^{-1}$	& \citet{Torres08}	\\
\noalign{\smallskip}
$\varv_{\rm sys}$ 	&  --14.7652\,(16)\,km\,s$^{-1}$	& \citet{Mazeh00}	\\
\noalign{\smallskip}
$a$		& 0.04707\,$^{+0.00046}_{-0.00047}$\,au	& \citet{Torres08} \\
\noalign{\smallskip}
$e\,\cos\omega$ &  0.00004\,(33)	& \citet{Crossfield12} \\
\noalign{\smallskip}
${P_{\rm orb}}$ & 3.52474859\,(38)\,d & \citet{Knutson07} \\
\noalign{\smallskip}
$T_{\rm 0}$ (HJD) & 2452826.628521\,(87)\,d & \citet{Knutson07} \\
\noalign{\smallskip}
{\em i} & $86.71(5)$\,deg  & \citet{Torres08} \\
\noalign{\smallskip}
$R_{\rm P}$$^{b}$ 	& 1.359\,$^{+0.016}_{-0.019}$\,R$_{\rm Jup}$	& \citet{Torres08}	\\
\noalign{\smallskip}
$K_{\rm P}$ &	$140(10)$\,km\,s$^{-1}$ 	& \citet{Snellen10}	\\
\noalign{\smallskip}
$\varv_{\rm wind}$ 	&  --5.2\,$^{+2.6}_{-1.3}$\,km\,s$^{-1}$	&	This work\\
\noalign{\smallskip}
\hline
\end{tabular}
\tablefoot{
The numbers in parentheses in the middle column correspond to the error bar on the last significant digits.
\tablefoottext{a}{\citet{Gaia18}.}
\tablefoottext{b}{Equatorial radius.}}
\end{table}

\section{Data reduction}
\label{datareduction}

\begin{figure}
\includegraphics[angle=0, width=1.0\columnwidth]{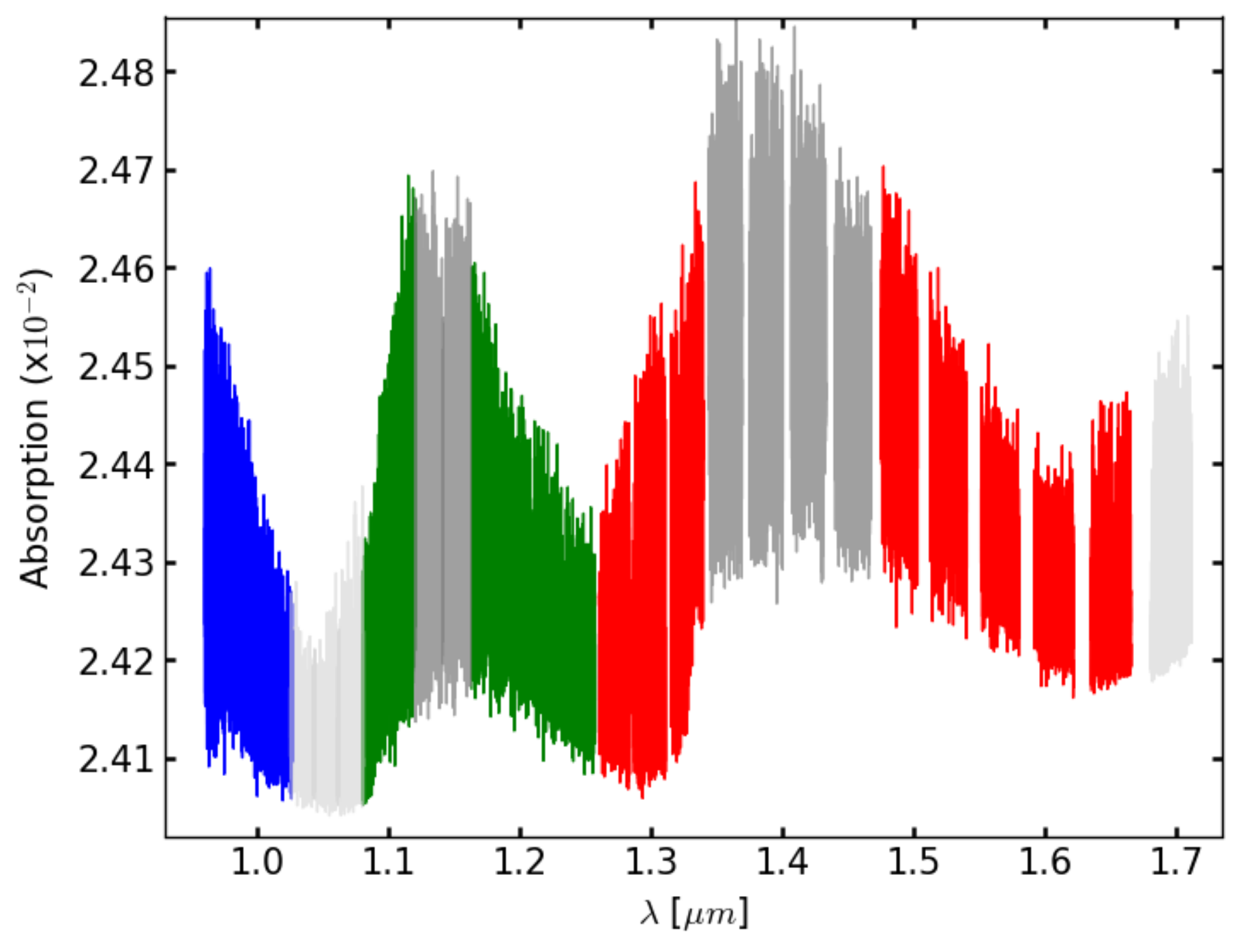}\hspace*{0.35cm}
\caption{Water vapor model transmission spectra computed for the best-fitting $p$-$T$ profile in \cite{Brogi17} and an \h2o volume mixing ratio of 10$^{-5}$. The model is shown at the same wavelengths of the data. The CARMENES line spread function for the NIR channel has been applied. The 1.0\,$\mu$m, 1.15\,$\mu$m and 1.4\,$\mu$m bands are shown in blue, green, and red, respectively. The orders in dark gray represent the a priori masks. The orders in light gray were masked a posteriori.}
\label{fig.syn_model}
\end{figure}

The raw spectra were processed by the CARMENES data reduction pipeline, {\sc Caracal} v2.20 \citep{Zechmeister2014, Caballero2016}.
As in \cite{Alonso19}, we used the European Southern Observatory {\sc Molecfit} tool \citep{Smette2015} to retrieve the column depth of precipitable water vapor (PWV) towards the target, which dropped from 10.3\,mm to 3.8\,mm during the observations (see Fig.\,\ref{pwv_airmass_sn}, top panel). The mean value of the PWV was $\sim$\,5.8\,mm, being the average PWV over Calar Alto of $\sim$\,7\,mm. Even with the relatively low mean PWV, there was a strong telluric absorption in the centers of the two strongest \h2o bands, which resulted in a very low flux.
We thus excluded from the analysis the 1.12--1.16\,$\mu$m (orders 54--53) and 1.34--1.47\,$\mu$m (orders 45--42) spectral regions (see the a priori masks in dark gray in Fig.\,\ref{fig.syn_model}). 

Three water vapor bands are observable with the NIR channel of CARMENES, each one covered by a different number of orders. Firstly, the spectral region 0.96--1.05\,$\mu$m (the 1.0\,$\mu$m band from now on) was covered by five useful spectral orders (63--59). Secondly, the spectral interval at 1.05--1.26\,$\mu$m (1.15\,$\mu$m band in the following) was covered by eight useful orders (58--55, 52--49). And, thirdly, the spectral region of 1.26--1.71\,$\mu$m (the 1.4\,$\mu$m band in the following), which was covered by nine spectral orders (48--36). In addition to the orders initially discarded by the a priori masks, we found orders 59\,--\,57 (1.03--1.08\,$\mu$m) and 36 (1.68--1.71\,$\mu$m) to present strong uncorrected telluric residuals in the analysis presented in Sect.\,\ref{crosscorr}. Therefore, we also discarded them in the analysis (see the a posteriori masks in Fig.\,\ref{fig.syn_model}). The 1.0\,$\mu$m, 1.15\,$\mu$m and 1.4\,$\mu$m bands were then covered by four, six, and eight useful orders, respectively.

Next, we removed from the spectra the 5\,$\sigma$ outliers, which were likely produced by cosmic rays. To do so, we fit an outlier-resistant third-order polynomial to the time evolution of the pixel and assigned the value of the fit to the affected frame. In addition, we derived the instrument drift during the night by measuring the shifts of the strong telluric lines \cite[as in][]{Alonso19}, and found a value of $\sim$\,0.023 pixel ($\sim$\,30\,m\,s$^{-1}$). If large, sub-pixel drifts have been found to impact the data analysis of this technique. For instance, this was the case of \citet{Brogi13} and \citet{Birkby17}, who found $\sim$0.7 pixel ($\sim$1 kms$^{-1}$) drifts for detector 1 in CRIRES. Thus, the authors performed a correction by aligning spectra to a common wavelength solution. As discussed in \citet{Brogi16}, this correction had an error of 0.03 pixel ($\sim$\,50\,ms$^{-1}$) in the absence of noise. Our drift is roughly two orders of magnitude lower than what was reported for detector 1 in CRIRES and, also, lower than the ideal alignment error computed in previous works. Therefore, we do not expect a drift correction to significantly affect our results.

The resulting spectra were stored as 18 matrices (one per order) of dimension 68\,$\times$\,4080, i.e., the number of spectra times the number of wavelength bins of each order. This allowed us to analyze each order independently by following the methodologies presented in previous publications \citep{Brogi16, Brogi18, Birkby17, Hawker2018, Alonso19} and discussed below. 

\subsection{Telluric and stellar signal removal with \sc{Sysrem}} \label{tellurics}

\begin{figure*}
\centering
\includegraphics[angle=0,,scale=0.34]{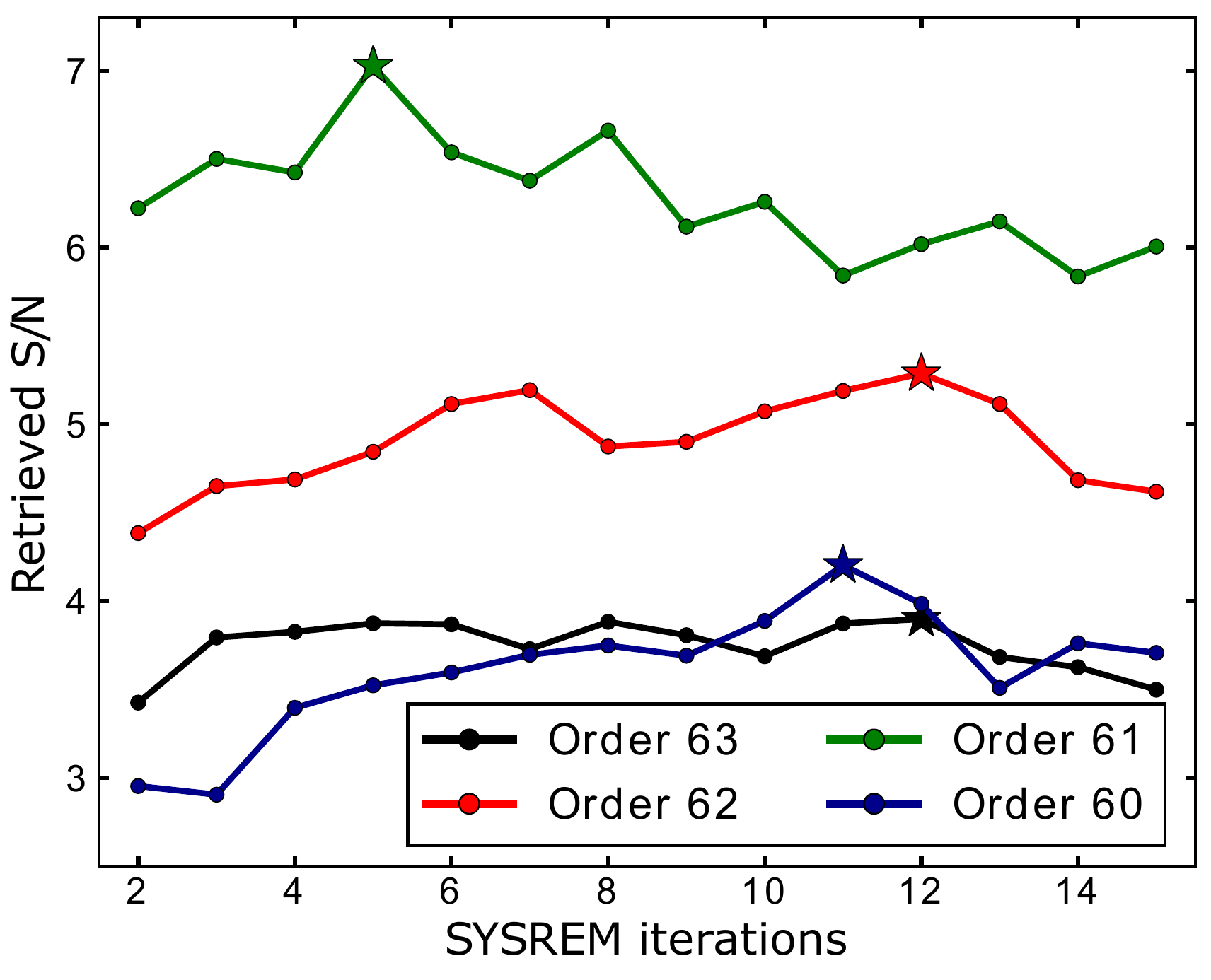}\includegraphics[angle=0,,scale=0.34]{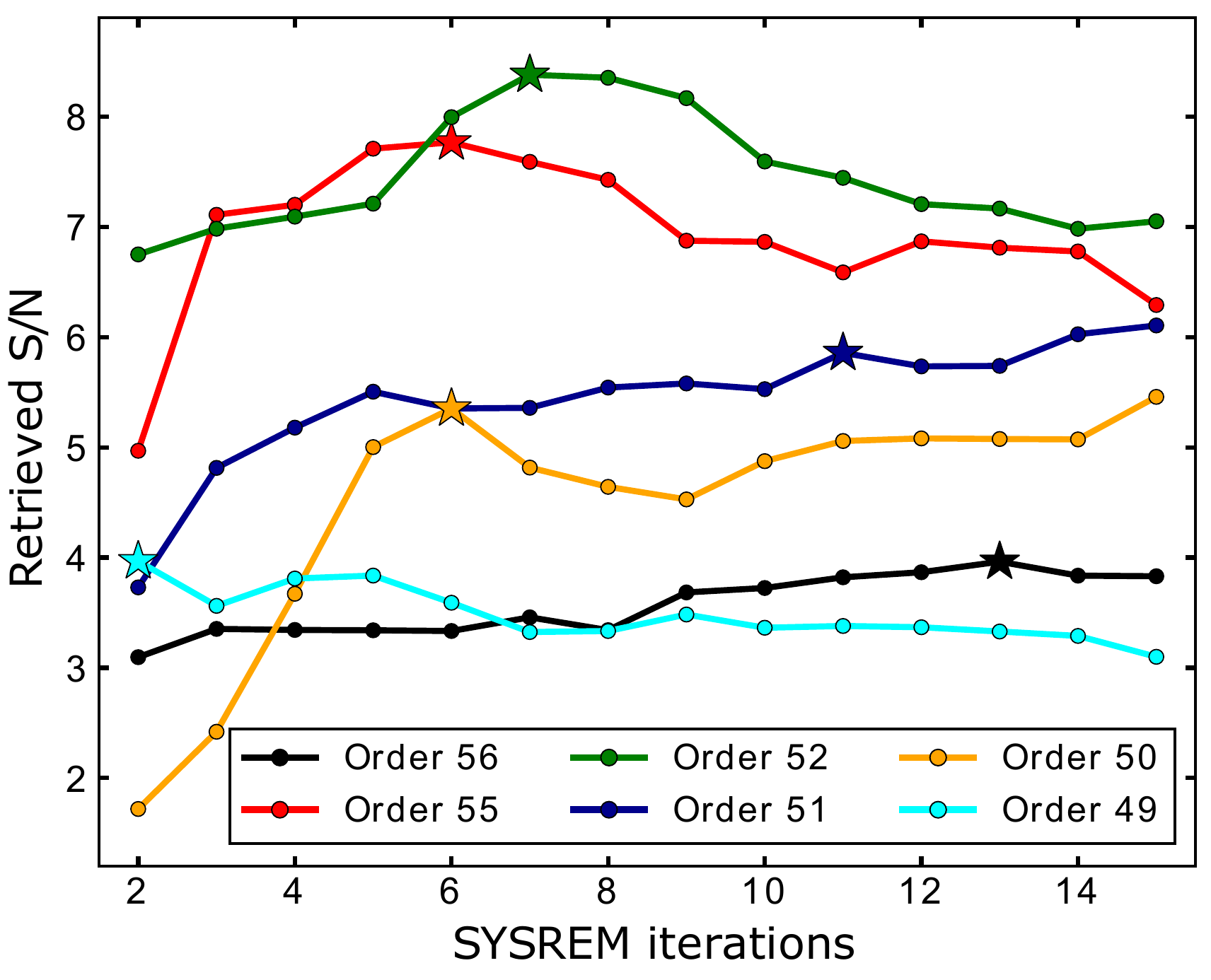}\includegraphics[angle=0,,scale=0.34]{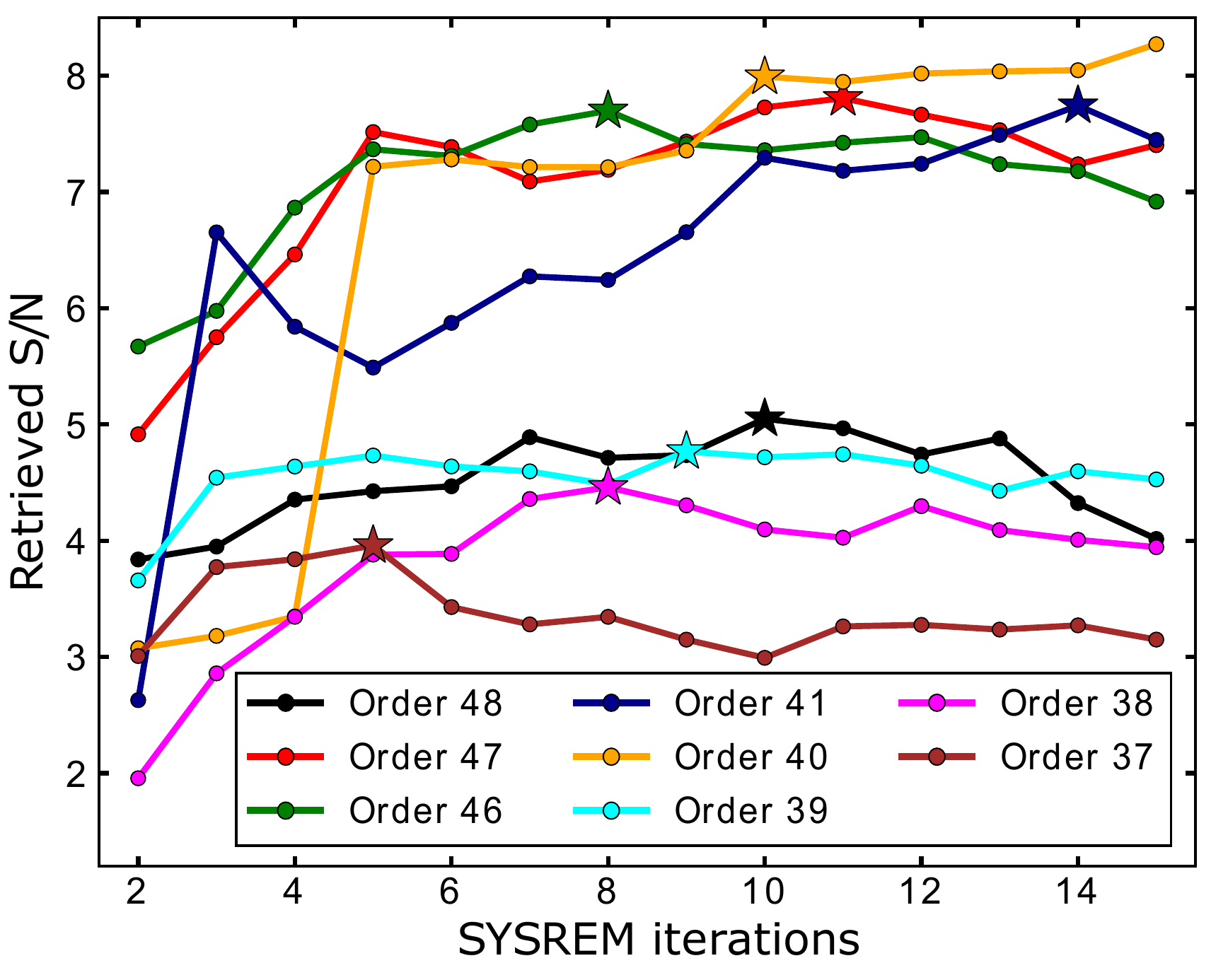}

\caption{Evolution of the retrieved S/N of the injected signal with the number of {\sc Sysrem} iterations for the 1.0\,$\mu$m band (\textit{left}), 1.15\,$\mu$m band (\textit{middle}), and 1.4\,$\mu$m band (\textit{right}). The model is injected at 5$\times$ the nominal strength. For the spectral orders where a very small signal is retrieved (i.e., S/N < 3), we injected a stronger model at 10$\times$ (for orders 49 and 38) and at 12$\times$ (for order 60) the nominal strength so as to ensure the injected signal is clearly measured (i.e., S/N $\geq$ 3). The stars represent the selected number of iterations for each spectral order at the peak of the recovered S/N (see text).}
\label{evol_bands}
\end{figure*}

Changing conditions in the Earth's atmosphere during the night (e.g., airmass variations) induce a variable pseudo-continuum level in the observed spectra (see Fig.\,\ref{sn_spectra}). In order to have the same baseline in all spectra, we normalized them order by order by fitting a quadratic polynomial to the pseudo-continuum. In this process, we excluded from the fit the spectral points corresponding to the sky emission lines identified by using fiber B, which are common in the reddest orders (see Fig.\,\ref{sn_spectra}).

Next, we masked the spectra at the wavelengths where the Earth's atmospheric absorption was larger than 80\% of the flux continuum. Thereby, we eliminated the spectral ranges where there is almost no flux. 
In our previous work \citep{Alonso19} we searched for the wavelength bins that satisfied the previous condition in the largest S/N spectrum of the night. This was a useful approach since the column depth of PWV during that night was stable and airmass did not change very much during the observations.
However, we found this procedure not to be suitable for the night analyzed here due to the rather large changes in the PWV, airmass, and S/N of the observed spectra (see Fig.\,\ref{pwv_airmass_sn}). For this reason, instead of selecting the spectrum with the largest S/N for masking, we chose that taken at the largest telluric water vapor absorption, which is more conservative. The masked regions represented $\sim$6\% of the data of the spectral orders included in the analysis.

The normalized and masked 68\,$\times$\,4080 matrices of spectra were at this point dominated by the quasi-static telluric and stellar contributions. The Doppler-shifted excess absorption from the planetary atmosphere is expected to be orders of magnitude lower than the telluric and stellar absorptions.
We removed these telluric and stellar signals by using {\sc Sysrem} (\citealt{Tamuz05, Mazeh07}), a principal component analysis algorithm that has been successfully applied to detect molecular signatures in the atmosphere of several exoplanets \citep{Birkby17, Nugroho17, Hawker2018, Alonso19}. This algorithm treats the temporal evolution of each wavelength bin as a light curve and removes systematic features common to all of them, while allowing each wavelength bin to be weighted by its uncertainty. In particular, the CARMENES pipeline provides a measurement of the photon noise and the read out noise for each pixel at each spectrum, which constitutes the input uncertainty matrix. By iterating {\sc Sysrem} in each order individually, we removed most of the stellar and telluric features in the spectra and also the variations induced by airmass and PWV changes. As was pointed out previously \citep[see, e.g.,][]{Nugroho17}, {\sc Sysrem} does not really fit airmass nor other atmospheric parameters. Instead, this algorithm performs linear fits that only approximate the non-linear evolution of the telluric absorption with time. Because of this, some telluric residuals are still expected to be left after each iteration. The planetary features also create a low-order effect at the sub-pixel level \citep{Birkby17}. Thus, after a certain number of {\sc Sysrem} iterations, these features are perceptible by the algorithm. At this point, {\sc Sysrem} starts focusing on fitting and removing the planetary signal, which is an unwanted effect. Also, as the intensity of the telluric and stellar contamination varies from one order to another, the optimum number of {\sc Sysrem} iterations is also expected to differ. This has been widely examined in the past \citep{Birkby17, Nugroho17, Hawker2018, Alonso19}.
Following these earlier studies, we analyzed the effect of {\sc Sysrem} order by order by injecting a model planetary signal (computed as described in Sect.\,\ref{models}) into the observations before removing the telluric contamination. In order to mimic the real signal of the planet, we Doppler-shifted the model planetary signal according to the expected orbital velocities of the planet at the times of the observed spectra. In other words, we injected the model at the expected velocities given by
\begin{equation}
\label{equation.planet_velocity}
\varv_{\rm P}(t,K_{\rm P}) = \varv_{\rm sys} + \varv_{\rm bary}(t)  + K_{\rm P}\sin{2\pi\phi(t)},
\end{equation}
where $\varv_{\rm sys}$ is the systemic velocity of the HD 209458 system, $\varv_{\rm bary}(t)$ is the barycentric velocity during the observations, $K_{\rm P}$ is the semi-amplitude of the orbital motion of the planet, and $\phi(t)$ is the orbital phase of the planet. In addition, possible atmospheric winds could be accounted for by considering an additional term, $\varv_{\rm wind}$, in the same manner as \cite{Alonso19}. 

We used the cross-correlation technique (see Secs.\,\ref{crosscorr} and \ref{significance}) to study the evolution of the S/N of the retrieved injected signal with the number of {\sc Sysrem} iterations. That is, we determined the optimum number of {\sc Sysrem} iterations for each spectral order by maximizing the significance of the injected cross-correlation function (CCF) peak \citep{Birkby17, Nugroho17, Hawker2018, Cabot19}. In order to characterize this behavior, we visually inspected the retrieved S/N evolution for each spectral order (see Fig.\,\ref{evol_bands}). 
The different performances can be broadly grouped into the following categories. Firstly, {\sc Sysrem} performed well for some spectral orders where a few iterations (< 10) were enough to reach the maximum recovery of the injected signal, and the significance then dropped (e.g., orders 61, 55, 52, 49, 46, 38, 37). For some spectral orders, the peak S/N was reached more slowly ($\geq$10 iterations, see, e.g., orders 60, 56, 48). Secondly, we observed a less efficient performance in some spectral orders where a maximum is achieved, but the behavior is almost flat (e.g., orders 63 and 39). In these cases, since the S/N dropped afterwards, we kept the found maximum. And thirdly, for some other orders the S/N improved slightly after the first maximum (e.g., orders 62, 51, 50, 47, 41, 40). In this latter case, we kept the iteration with the first maximum of the S/N if it did not improve significantly (changes smaller than 5\%) in subsequent iterations. We found this behavior for orders 62, 47, and 41.

We explored the removal of the telluric absorption by using {\sc Molecfit} instead of {\sc Sysrem}. However, we did not find any planetary signal when applying the procedure described in Sect.\,\ref{crosscorr} to the spectra obtained with this telluric correction. Recent works have applied this tool to CRIRES data and have found the scatter of the residuals to be between 3\% and 7\% \citep{Ulmer_Moll19}. Therefore, the fit uncertainties over the entire wavelength coverage of the CARMENES NIR channel seem to be too large for attempting a detection of the weak \h2o features of this planet.

\section{Planetary atmospheric signal extraction}
\label{signalretrieval}

We used the cross-correlation technique to extract the atmospheric signal. The basic idea is to accumulate the information of the thousands of absorbing ro-vibrational lines, which is usually expressed as the amplitude of a CCF. To do so, we computed high-resolution absorption models of the atmosphere for the primary transit that served as a template for the CCF. Water vapor (\h2o), methane (\ch4), ammonia (NH$_3$), hydrogen cyanide (HCN), and carbon monoxide (CO) are the main absorbent species in the wavelength range covered by the NIR channel of CARMENES for this planet. We have correlated the measured spectra with exo-atmospheric transmission models including the absorption of these molecules.

\subsection{High-resolution absorption models}\label{models}

We computed high-resolution spectral transmission models of \hd20 in its primary transit by using the Karlsruhe Optimized and Precise Radiative Transfer Algorithm \citep[{\sc Kopra};][]{Stiller2002}. This code was originally developed for the Earth's atmosphere and has been afterwards adapted to other planetary atmospheres in the Solar System \citep{Garcia-Comas2011, LopezPuertas2018}, and more recently also to exoplanet's atmospheres \citep{Alonso19}. {\sc Kopra} is a well-tested general purpose line-by-line radiative transfer model that includes all known relevant processes for computing those transmittances. 
In particular, we used a Voigt line shape for the ro-vibrational lines. The code uses an adaptive scheme for including or rejecting spectral lines at a specified strength and a given absorption accuracy \citep[see][]{Stiller2002}. This is particularly useful in our case given the large number of ro-vibrational lines of \h2o at high temperatures. The spectroscopic data were taken from the HITEMP 2010 compilation \citep{Rothman2010} for \h2o and from the HITRAN 2016 compilation \citep{Gordon17} for the other molecules.
Collision-induced absorption was included for H$_2$-H$_2$ and H$_2$-He \citep[see][for more details]{Alonso19}. Rayleigh scattering was included, although it is almost negligible in the NIR. We did not include clouds nor additional haze contributions.

We computed the transmission models at a very high resolution ($\mathcal{R}$\,$\sim$\,4$\cdot$10$^{7}$) and afterwards convolved them with the CARMENES NIR line spread function. We used the pressure-temperature profile retrieved by \cite{Brogi17} by combining low and high-resolution measurements for \hd20. In addition, we chose \h2o, \ch4, \nh3, HCN, and CO volume mixing ratios (VMR) of 10$^{-5}$, 10$^{-8}$, 10$^{-8}$, 10$^{-6}$, and 10$^{-4}$, respectively, which are very similar to the most probable values retrieved in that work. In Fig.\,\ref{fig.syn_model} we show the computed transmission model for water vapor where the major absorbing bands near 1, 1.15, and 1.4\,$\mu$m are identified. 

In the same way as was done for the observed spectra, we normalized the modeled transmission. Therefore, the study presented in the following is not sensitive to the absolute absorption level of the lines, but to their relative depth with respect to the continuum.

\subsection{Cross-correlation}
\label{crosscorr}

We cross-correlated the residual matrices with the transmission model (template) Doppler-shifted in the range of --200 to 200\,km\,s$^{-1}$ with respect to the Earth's rest-frame. The step size was 1.3\,km\,s$^{-1}$, which was calculated by averaging the velocity step-size of the pixels in the NIR channel. In each step, the template was linearly interpolated to the corresponding Doppler-shifted wavelength grid.
We obtained a CCF for each of the 68 spectra, thus forming cross-correlation matrices, $\overline{CCF}$s, of dimensions 201\,$\times$\,68.  These matrices were calculated independently for each spectral order and for each {\sc Sysrem} iteration. Additionally, we subtracted the median value from each row (i.e., 201 spectral velocity bins) of the matrices in order to remove broad variations arising from the differences between the spectra and the template.

\begin{figure}
\includegraphics[angle=0, width=1.0\columnwidth]{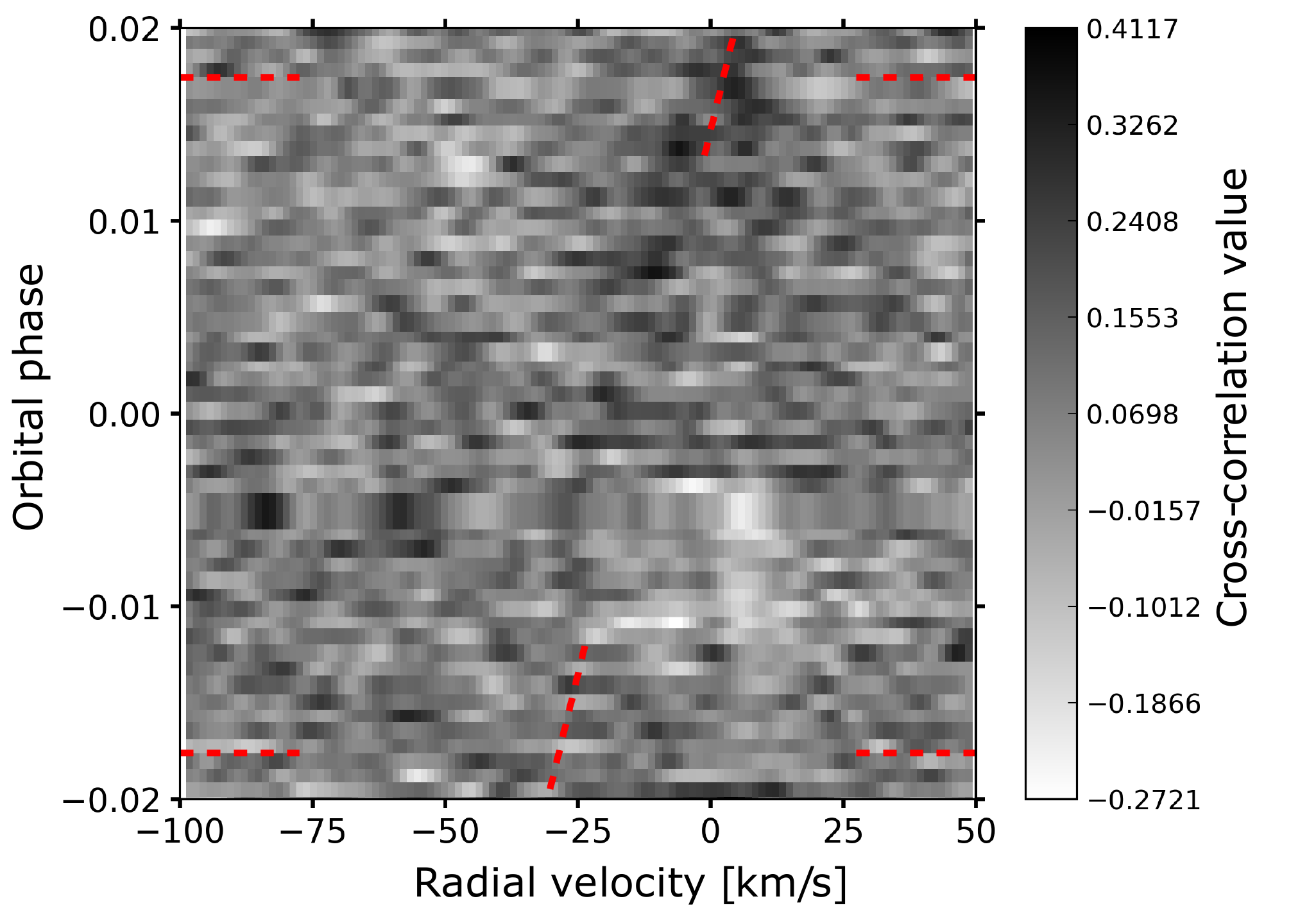}
\caption{Cross-correlation values as a function of the orbital phase and planet orbital velocity in the Earth's rest-frame. The results were obtained from 18 useful NIR orders. The orbital phase of the first and last in-transit spectra are indicated by horizontal dashed red lines. A visual guide to the expected exoplanet velocities (i.e., \kp\,=\,140\,km\,s$^{-1}$ and assuming no additional winds in the atmosphere of the planet) is indicated with tilted dashed red lines.} 
\label{ccf_matrix_earth}
\end{figure}

We further enhanced the signal by co-adding the information in the $\overline{CCF}$s from the different spectral orders. The resulting total $\overline{CCF}$ in the Earth's rest-frame is shown in Fig.\,\ref{ccf_matrix_earth}. The trace of the planet is expected to appear during transit (i.e., between the horizontal dashed lines) as positive cross-correlation values along the planetary velocities $\varv_{\rm P}$ calculated by Eq.~\ref{equation.planet_velocity}. During the observing night, these velocities were expected to range roughly between $-$34 and 15\,km\,s$^{-1}$ (see tilted dashed lines).

Next, we aligned the rows of the cross-correlation matrices to the rest frame of the planet. In this process we assumed \kp\ in Eq. \ref{equation.planet_velocity} as unknown and performed the shift for a \kp\ grid from --280 to 280\,km\,s$^{-1}$. As in earlier studies, this allowed us to directly measure K$_{\rm P}$, although with rather large uncertainties. This is mainly because the change in the exoplanet radial velocity during the transit, of a few tens of \kms, is too small to accurately measure it with this technique \citep{Brogi18, Alonso19}. Also, we were able to check for strong telluric residuals that might appear close to the expected exoplanet velocities or around zero or negative \kp\ values.
Consecutively, we co-added the cross-correlation values over time and obtained the total CCF. We excluded from the co-adding the last seven in-transit spectra at 0.012\,<\,$\phi$\,< 0.018, for which the velocity of the planet with respect to the Earth was very small (i.e., between $\pm$\,2.5\,\kms) and hence, the absorption lines from the planetary atmosphere nearly overlapped with the telluric lines. At this point, if the atmospheric molecular signal were strong enough, we would expect to find a CCF peak around the zero velocity and the expected \kp\,=\,140\,km\,s$^{-1}$. 

\begin{figure}
\includegraphics[angle=0, width=1.0\columnwidth]{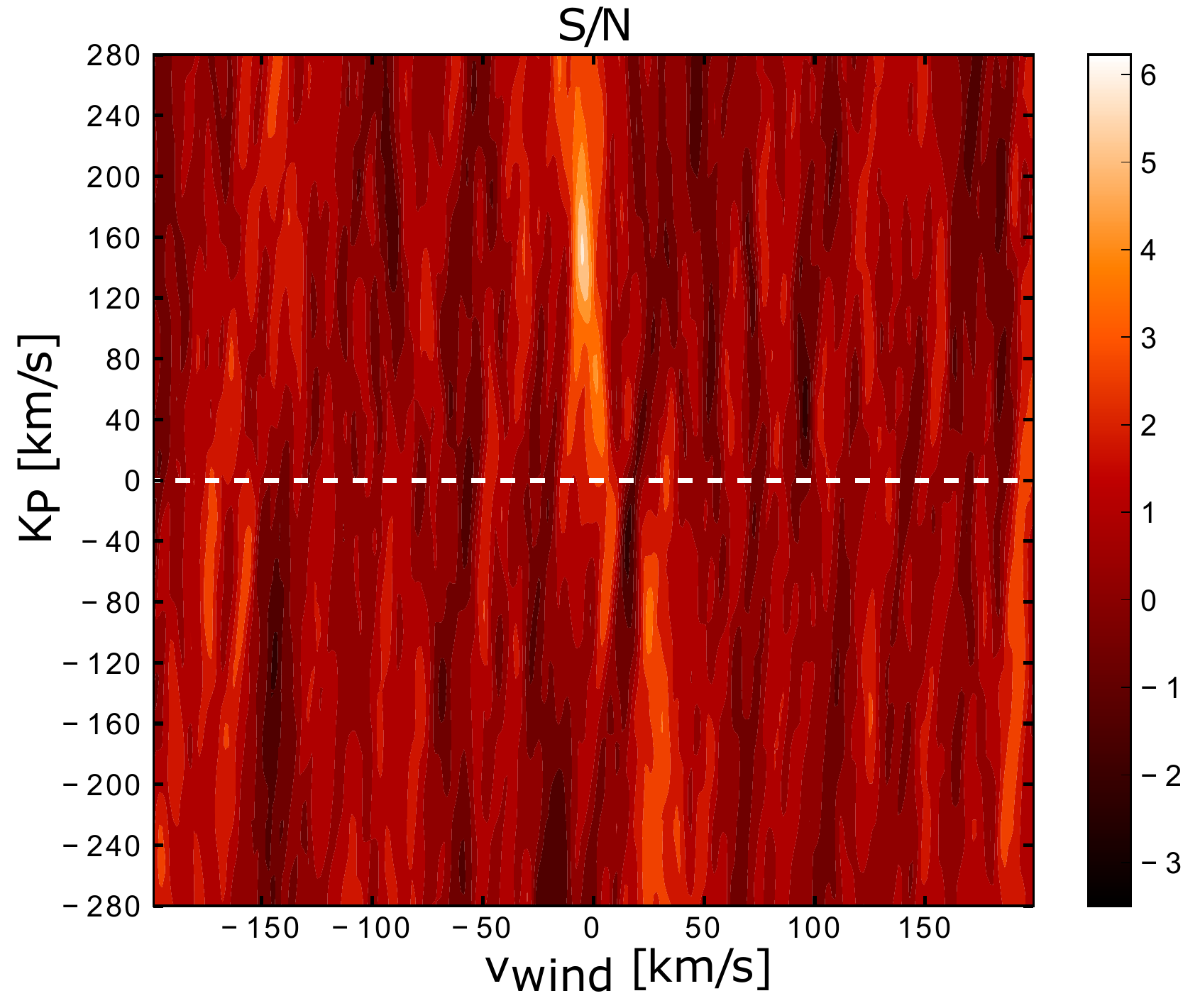}
\includegraphics[angle=0, width=1.0\columnwidth]{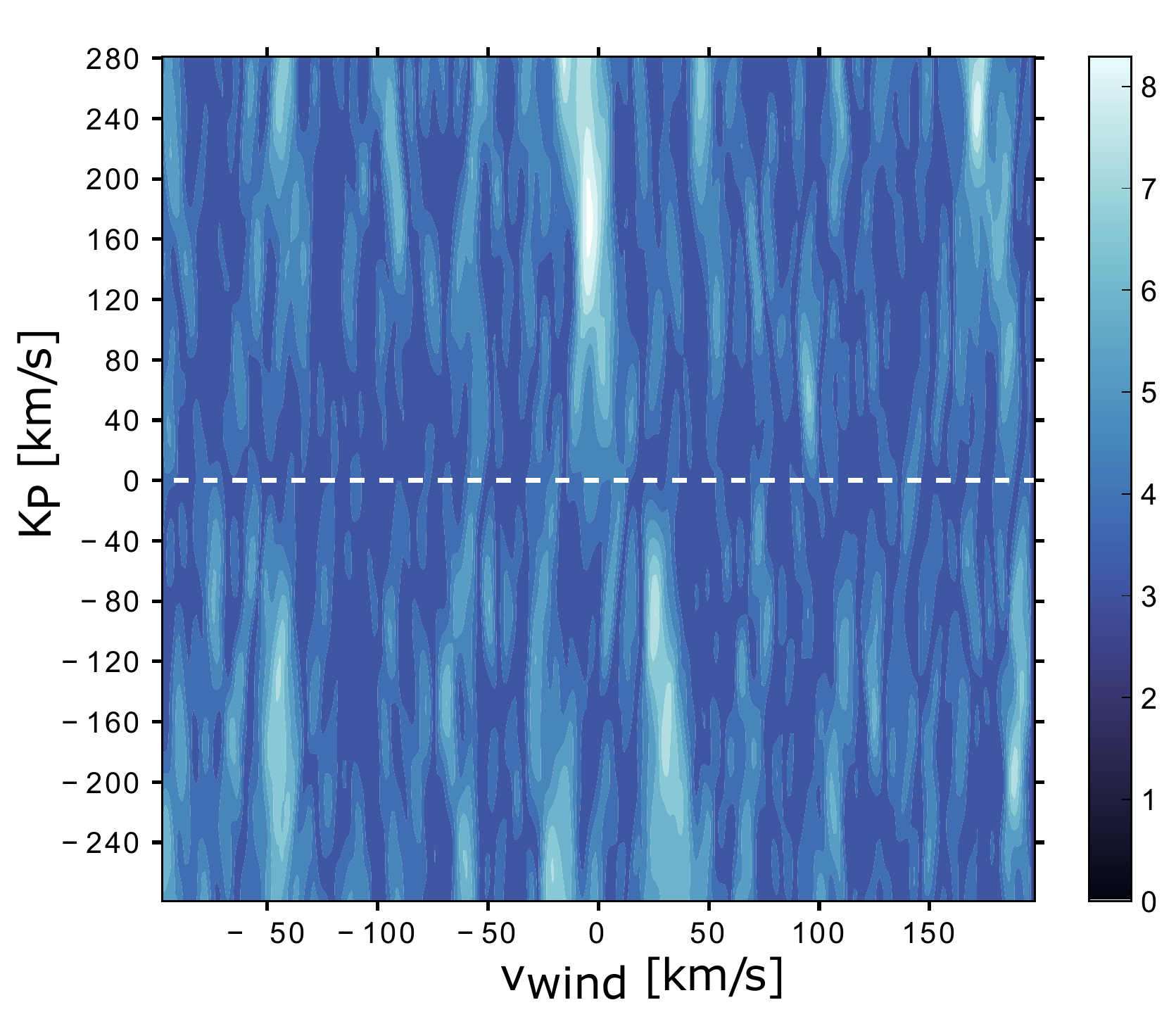}
\caption{S/N map (\textit{top}) and  $\sigma$ map  (\textit{bottom}) obtained after the cross-correlation of the residual spectral matrices with the atmospheric transmission template. All useful NIR orders were included. A dashed white line marks the \kp\,=\,0\,\kms value. Both maps show the region of maximum significance at very similar \kp\ values ($\sim$\,150\,\kms).}
\label{ccf_matrices} 
\end{figure}

\subsection{Measurement of the significance of the signal}
\label{significance}

We obtained the significance of the signal by applying the same methodology as in \citet{Birkby17}, \citet{Brogi18}, and \citet{Alonso19}. First, we computed a S/N map (see top panel in Fig.~\ref{ccf_matrices}), where the S/N values were calculated at each \kp\ by dividing the CCF values by the CCF standard deviation obtained from the whole velocity interval but excluding the  $\pm$\,15.6\,km\,s$^{-1}$ interval around the CCF value. As a result, the S/N thus depends on the width of the velocity interval. The presence of a planetary signal in the map should appear then as a region of large S/N at a \kp\ compatible with that of \hd20, which is observed in the top panel of Fig.~\ref{ccf_matrices} at around \kp\ $=$\,150\,km\,s$^{-1}$.

We further examined the significance of the signal by using the in-transit cross-correlation matrix in the rest-frame of the planet. Here we defined two distributions: the ``in-trail" distribution, covering the velocities from --1.3\,km\,s$^{-1}$ to 1.3\,km\,s$^{-1}$ (i.e., three pixels), where we expected the exoplanet signal, and the ``out-of-trail" distribution comprising the rest of the velocities but excluding the $\pm$\,15.6\,km\,s$^{-1}$ region. The out-of-trail-distribution should contain non-correlated noise, hence being normally distributed with zero mean.
We compared the two distributions by performing a Welch's $t$-test \citep{Welch47}, following earlier studies \citep{Birkby17, Alonso19, Cabot19}. To do so, we set a null hypothesis, H$_0$, of the in- and out-of-trail samples having the same mean. The test returns the value of the $t$--statistic, which is expected to be larger the stronger the evidence against H$_0$ is. Each $t$-value can be directly converted into a $p-$value that represents the probability of obtaining a $p$-value as large as observed or larger if H$_0$ was true. Subsequently, we expressed the $p$-values in terms of standard deviations ($\sigma$-values). In this analysis, the presence of a signal of atmospheric origin in our data would yield significantly different means for the two distributions. The in-trail cross-correlation values would then contain information not consistent with uncorrelated noise, hence rejecting the null hypothesis.

\begin{figure}
\includegraphics[angle=0, width=1.0\columnwidth]{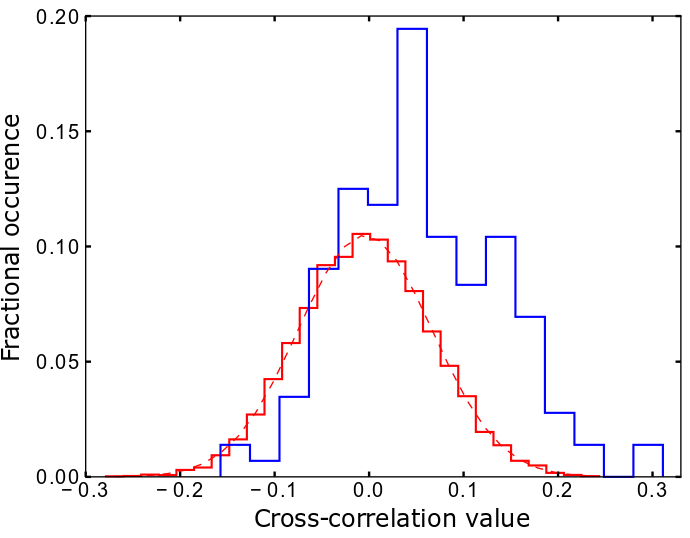}
\caption{Distribution of cross-correlation values around the expected planetary radial velocities (in-trail, blue), far away from the planet's velocities (out-of-trail, red), and a Gaussian distribution with the same mean and variance as the out-of-trail distribution (dashed red). As expected for a signal of planetary origin, the in- and out-of-trail distributions show significantly different means.
}
\label{distributions}
\end{figure}

In Fig.~\ref{distributions} we show both distributions for the pairs of (\kp,\,$\varv_{\rm wind}$) that yield the largest $\sigma$-value. The means of the in-trail and out-of-trail distributions are different, with the latter following a Gaussian distribution. Thus, this test also provides evidence for the presence of a planetary signal in our transit data. In order to better characterize this signal, we computed a map of $\sigma$-values for the same \kp\ grid (see bottom panel of Fig.\,\ref{ccf_matrices}) and found a region of maximum $\sigma$-values consistent with the S/N results (top panel).

\section{Results and discussion}
\label{results_discussion}

\begin{figure}
\includegraphics[angle=0, width=1.0\columnwidth]{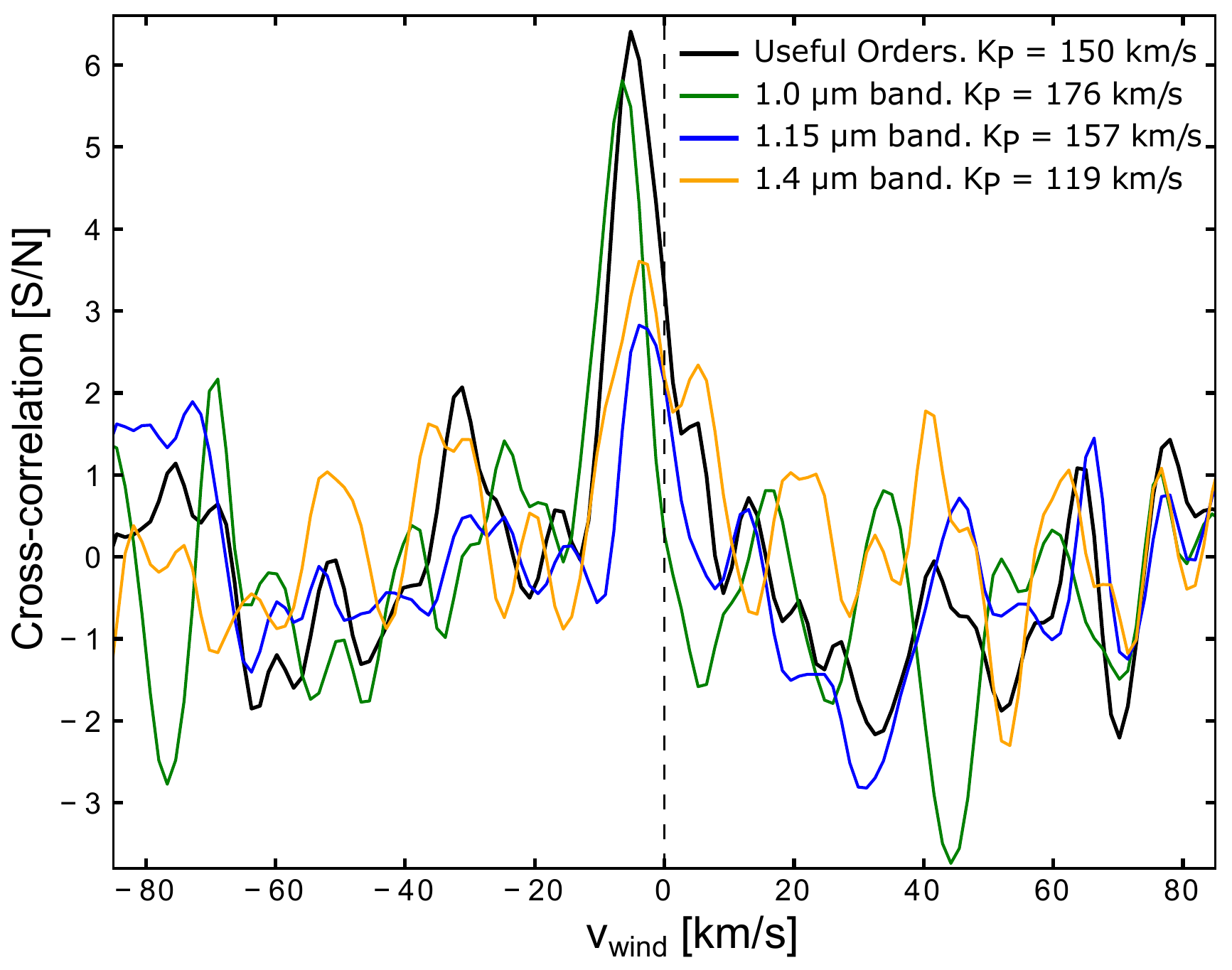}
\caption{CCFs with the largest significance obtained for the 1.0\,$\mu$m band (green), 1.15\,$\mu$m band (blue), and including all useful NIR orders (black). Also, the CCF for the 1.4\,$\mu$m band with the largest significance peak in the positive \kp\ space is shown (orange). The vertical dashed line marks the zero velocity value.
}
\label{ccf_weight_diff_it}
\end{figure}

We have studied the detectability of water vapor in \hd20 transit spectra with the NIR channel of CARMENES. We computed the total CCF when including all the useful NIR spectral orders (see black curve in Fig.\,\ref{ccf_weight_diff_it}) and detected \h2o with a maximum S/N\,=\,6.4 and a $\sigma$-value of 8.1, which was larger than its S/N. Similar results were found by \cite{Cabot19}, indicating that the Welch's $t$-test might overestimate the significance of the signal more than the S/N calculation. However, the noise in the S/N calculations might contain contributions from the auto-correlation function of the template (i.e., aliases) or, in general, correlated noise that still averages to zero, but inflates the standard deviation. In this case, the S/N calculation might be an underestimation of the potential signal in the data.

The S/N and the $\sigma$ maps for the explored \kp\ grid are depicted in the top and bottom panels of Fig.\,\ref{ccf_matrices}, respectively.
We found the maximum signal at a \kp\,=\,150\,$^{+28}_{-25}$\,km\,s$^{-1}$, which is in line with the value derived by \cite{Hawker2018} from CRIRES observations, within the uncertainties. The cross-correlation with models including \ch4, \nh3, HCN, or CO individually did not yield any CCF peak compatible with a planetary origin. Consequently, the models including \h2o+\ch4, \h2o+\nh3, \h2o+HCN, or \h2o+CO showed noisier CCFs, and lower S/N than that obtained by including water vapor alone. However, the strongest absorption features of \ch4, \nh3, HCN, and CO occur at the reddest NIR orders of CARMENES, which presented the lowest S/N and the least efficient telluric removal in our data. Therefore, our results for these molecules are not conclusive and should be validated with future observations.

\begin{figure*}
\includegraphics[angle=0,scale=0.36]{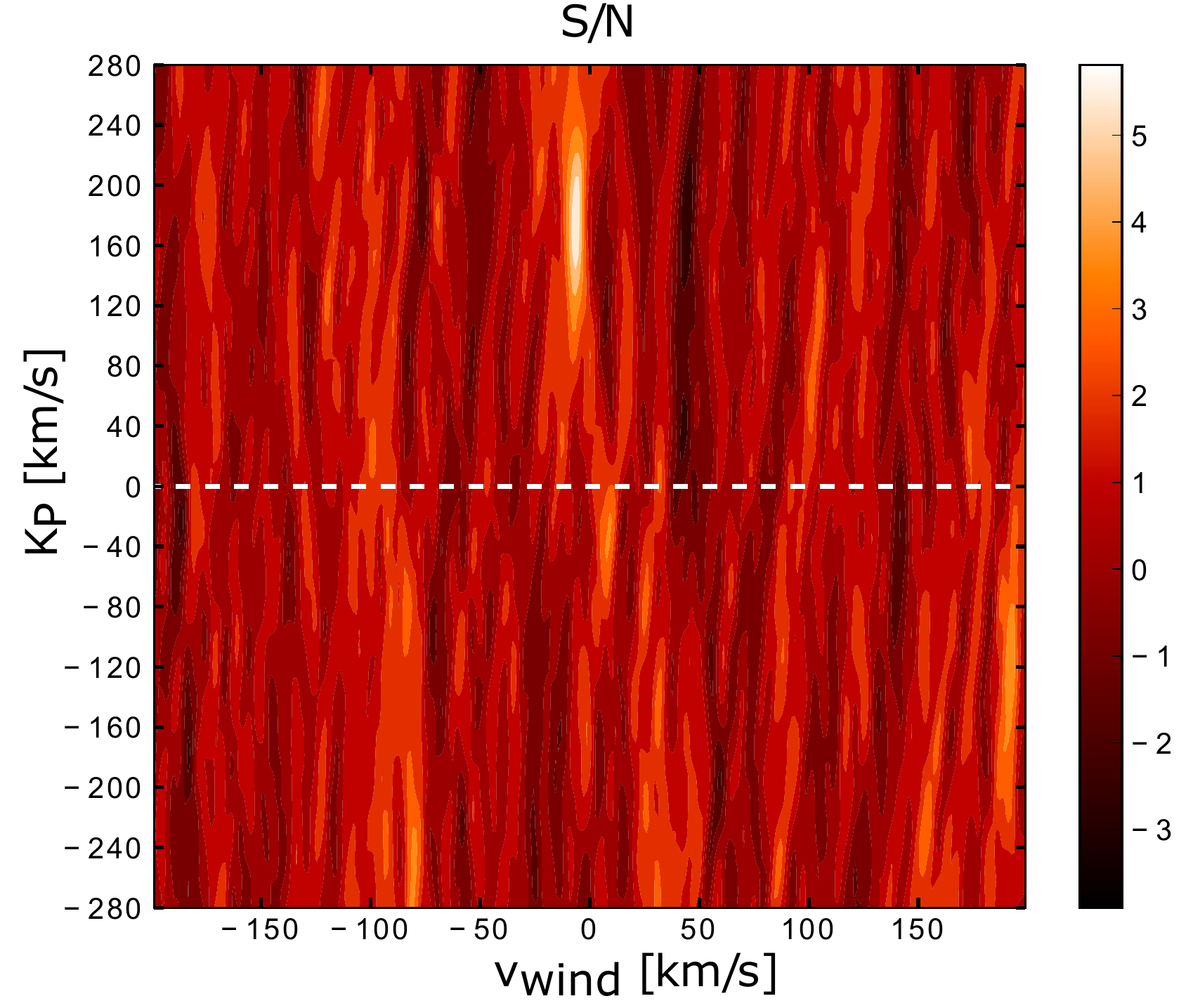}\includegraphics[angle=0,scale=0.36]{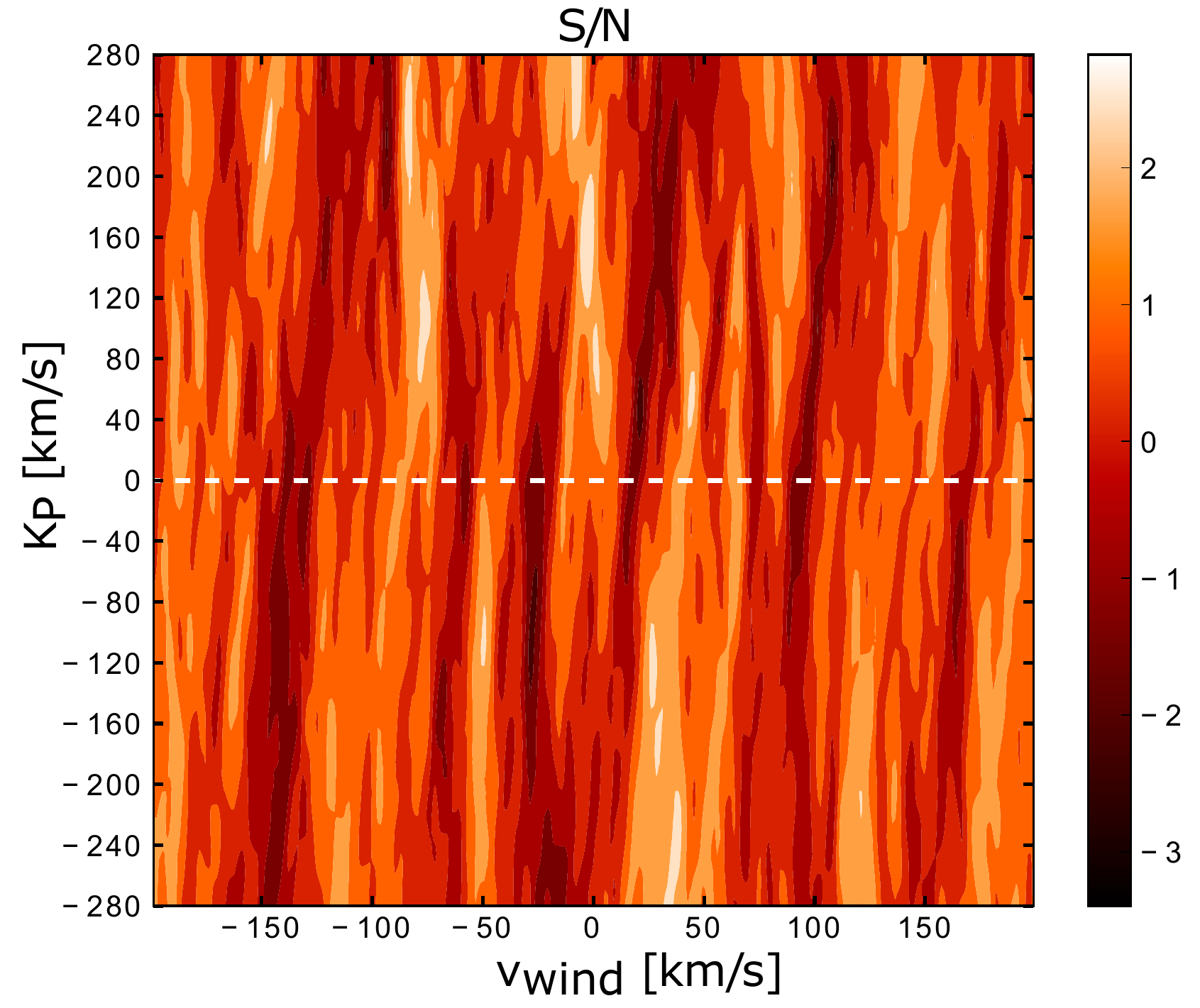}\includegraphics[angle=0,scale=0.36]{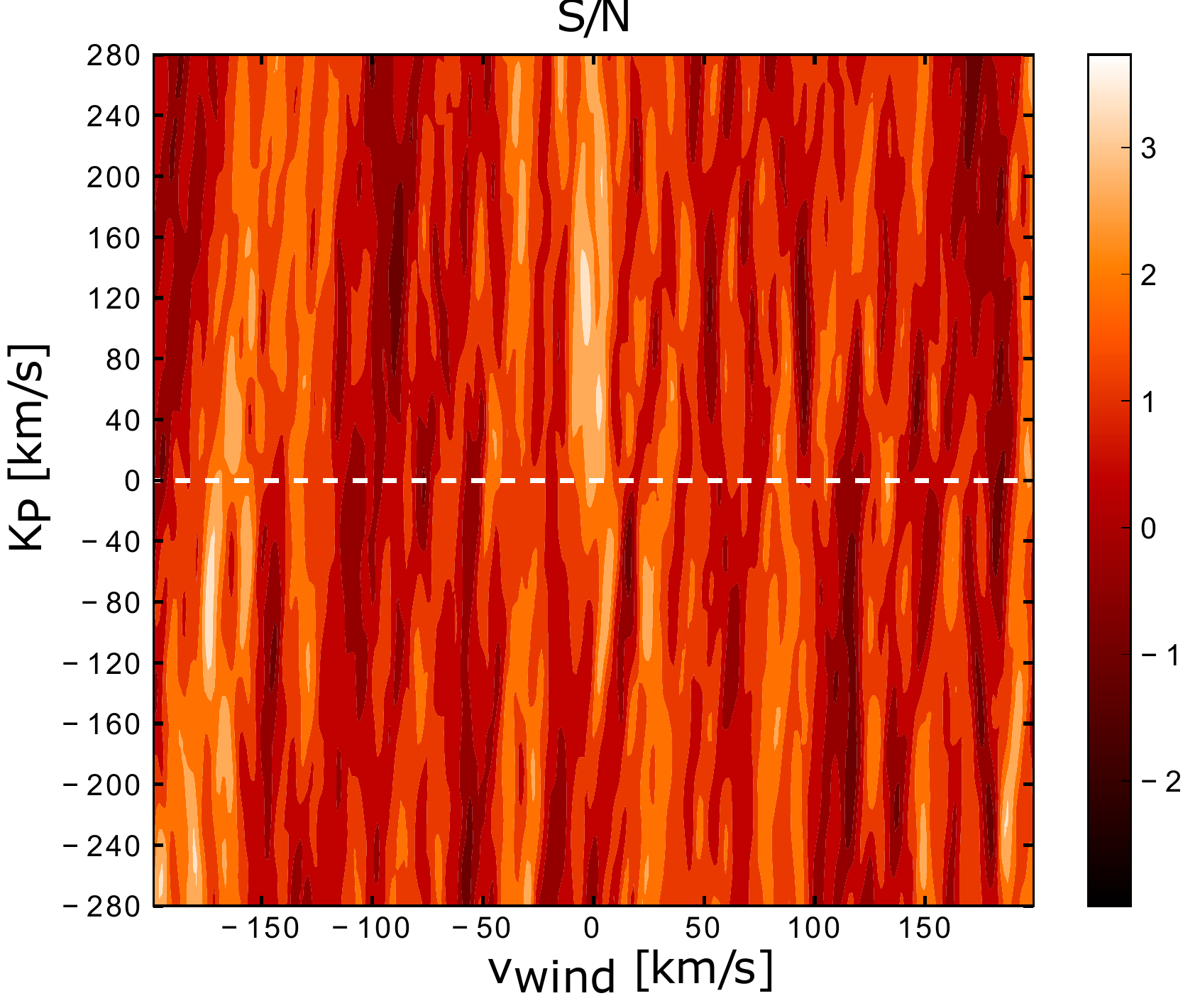} 
\caption{S/N maps obtained for \hd20 after the cross-correlation of the residual spectral matrices with the atmospheric transmission template for the 1.0\,$\mu$m band (left), 1.15\,$\mu$m band (middle) and for the 1.4\,$\mu$m band (right). A dashed white line marks the \kp\,=\,0\,\kms\ value.}
\label{multiband_maps}
\end{figure*}

\subsection{Multi-band analysis}
\label{multiband}

We have investigated the possibility of detecting water vapor by using separately the data in the three bands covered by the CARMENES NIR channel (see Figs.\,\ref{fig.syn_model} and \ref{multiband_maps} and Table\,\ref{table.global_results}). We detected \h2o from the 1.0\,$\mu$m band with a maximum S/N of 5.8 ($\sigma$-value of 7.4) at a \kp\,=\,176\,$^{+30}_{-38}$\,km\,s$^{-1}$ (see green curve in Fig.\,\ref{ccf_weight_diff_it} and left panels in Fig.\,\ref{multiband_maps}). However, \h2o was not detected using this band, covered only partially by the NIR channel, in \hdu18 \citep{Alonso19}. Actually the detection in \hd20 represents the first detection of \h2o from this band in any exoplanet so far. 

Regarding the 1.15\,$\mu$m band, we found the S/N map to be noisier than for the 1.0\,$\mu$m band, possibly due to a less efficient telluric correction in this spectral region (see middle panel in Fig.\,\ref{multiband_maps}). We obtained the maximum significance CCF peak in the map with S/N of 2.8 ($\sigma$-value of 6.9) at a \kp\,=\,157\,$^{+33}_{-47}$\,km\,s$^{-1}$ (see blue curve in Fig.\,\ref{ccf_weight_diff_it}), which is consistent with the expected velocity of the planet. However, this S/N is rather low and some contamination can be seen in the form of other similar peaks. In particular, the contamination at the negative \kp\ space is carried over when including this band in all the useful NIR orders (see negative \kp\ space in the S/N and $\sigma$ maps in Fig.\,\ref{ccf_matrices}). Therefore, the maximum CCF peak of the 1.15\,$\mu$m band should be interpreted only as a hint of a signal, since the telluric removal in these orders was less efficient. The results are worse for the 1.4\,$\mu$m band, for which different maxima appear (see right panel in Fig.\,\ref{multiband_maps}), being the maximum significance peak of the map located at the negative \kp\ space. The maximum significance CCF peak in the positive \kp\ space is found with a S/N\,=\,3.6 ($\sigma$-value of 5.3) at a \kp\,=\,119\,$^{+43}_{-45}$\,km\,s$^{-1}$ (see orange curve in Fig.\,\ref{ccf_weight_diff_it}), which is consistent with the expected value within the uncertainties. However, this result is largely insufficient for claiming a detection from this band individually.

Overall, these results indicate that most of our signal arose from the 1.0\,$\mu$m band, and partially from 1.15\,$\mu$m band. However, including this spectral region in the analysis we obtained a slightly larger signal. It is surprising though that we obtained the largest contribution to the signal from the weakest band rather than from the two strongest 1.15\,$\mu$m and 1.4\,$\mu$m bands. It is true that, for the strongest bands, some orders near the center of the bands were discarded because of the absence of flux. Additionally, the masking of the strongest absorption lines (i.e., absorbing more than 80\,\% of the flux) within each order rejects many spectral points and hence reduces the planetary signal. But this is in contrast with what we would expect \citep[see, e.g.,][]{Alonso19} where we obtained the largest signals in the stronger bands. Several reasons might explain this. Firstly, the mean S/N per pixel of the observed spectra in the second half of the observations fall rapidly (see Fig.\,\ref{pwv_airmass_sn}, bottom panel). In fact, the last spectra have around half of the S/N of the first ones. Additionally, the S/N within each spectrum decreases towards longer wavelengths with typical values in the range 20\,--\,60 at the reddest orders (see Fig.\,\ref{sn_spectra}). This was possibly due to the large airmasses in the second half of the observations, between 1.3 and 1.8 (Fig.\,\ref{pwv_airmass_sn}, middle panel). Secondly, the drastic drop of the column depth of PWV (Fig.\,\ref{pwv_airmass_sn}, top panel) that occurred during the observations possibly added other trends to the data that make more difficult the telluric removal. These effects combined might have hampered an efficient signal retrieval from the 1.15\,$\mu$m and 1.4\,$\mu$m bands.

\begin{table}
{\tiny
\centering
\caption{\label{table.global_results} Maximum S/N and $p-$values of the CCFs at the analyzed wavelength intervals for \hd20.}
\begin{tabular}{lcccc} 
\hline
\hline
\noalign{\smallskip}
Band & S/N & $\sigma$-values & $K_{\rm P}$ [km\,s$^{-1}$] & $\varv_{\rm wind}$ [km\,s$^{-1}$] \\
\noalign{\smallskip}
\hline
\noalign{\smallskip}
1.0\,$\mu$m & 5.8 & 7.4 & 176\,$^{+30}_{-38}$ & --6.5\,$^{+2.6}_{-1.3}$\\
  \noalign{\smallskip}
1.15\,$\mu$m & 2.8 & 6.9 & 157\,$^{+33}_{-47}$ & --3.9\,$^{+3.9}_{-2.6}$\\
 \noalign{\smallskip}
1.4\,$\mu$m$^{a}$ & 3.6 & 5.3 & 119\,$^{+43}_{-45}$ & --3.9\,$\pm$\,2.6 \\ 
\noalign{\smallskip}
Useful orders$^{b}$  & 6.4 & 8.1 & 150\,$^{+28}_{-25}$ & --5.2\,$^{+2.6}_{-1.3}$\\
\noalign{\smallskip}
\hline
\end{tabular}
\tablefoot{$^{a}$ Maximum significance CCF peak measured in the positive \kp\ space.
$^{b}$Combination of the 18 useful NIR spectral orders.}
}
\end{table}

\begin{figure*}
\centering
\includegraphics[angle=0,,scale=0.36]{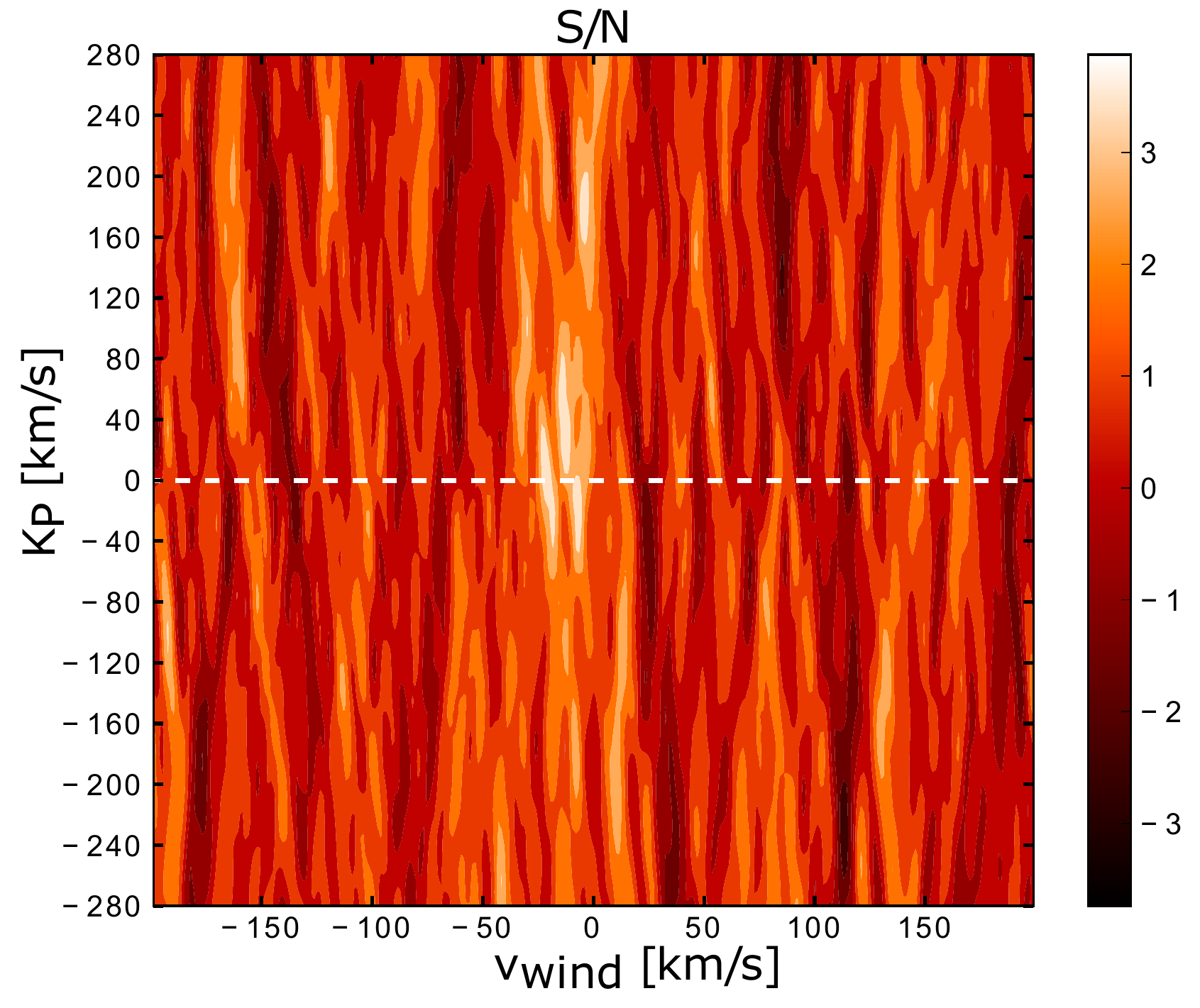}\includegraphics[angle=0,,scale=0.36]{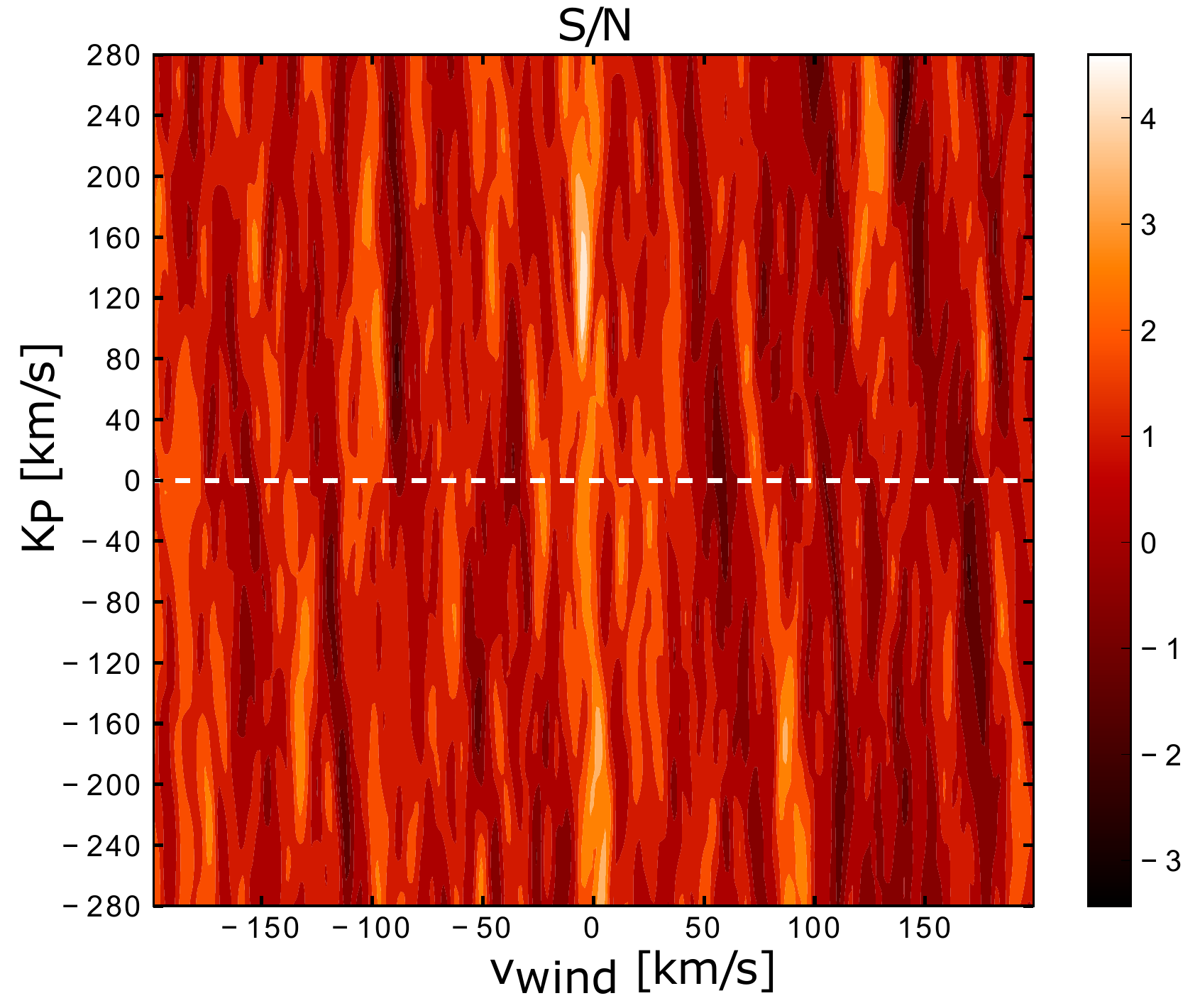}\includegraphics[angle=0,,scale=0.36]{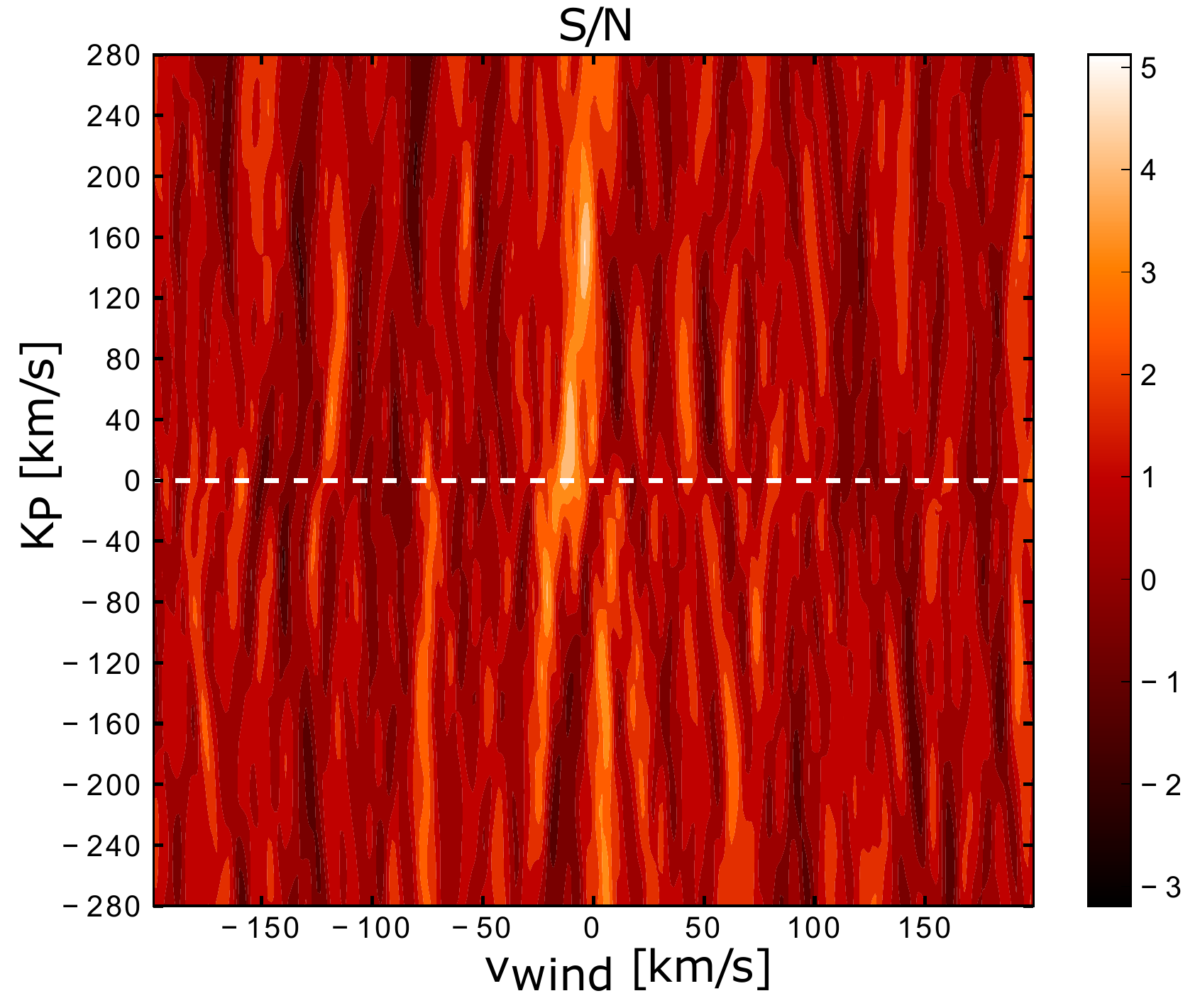}
\caption{S/N maps obtained for CARMENES observations of \hdu18 for the 1.0\,$\mu$m band (left panel), for the 1.15\,$\mu$m band (middle panel) and for the 1.4\,$\mu$m band (right panel). A dashed white line marks the \kp\,=\,0\,\kms\ value.}
\label{multiband_maps_hd18}
\end{figure*}

\subsection{Winds in the atmosphere of \hd20}
\label{winds}

The CCF peaks showed a net blueshift that was $-$\,6.5\,$^{+2.6}_{-1.3}$\,\kms\ for the 1\,$\mu$m band, $-$3.9\,$^{+3.9}_{-2.6}$\,\kms\ for the 1.15\,$\mu$m band, and $-$5.2\,$^{+2.6}_{-1.3}$\,\kms\ when considering all the useful NIR spectral orders. This could be caused by strong winds flowing from the day- to the night-side at the terminator of this hot Jupiter. 
The different absolute wind velocities retrieved from both bands could be related to the different pressure levels at which the bulk of the spectrum is formed for each wavelength interval. However, we would expect the weaker 1.0\,$\mu$m band to have its largest contribution at slightly higher pressure levels (i.e., lower altitudes) than the stronger 1.15\,$\mu$m band and hence, to have a lower $\varv_{\rm wind}$ according to global circulation models \cite[e.g., see Fig.\,3 in][]{Rauscher12, Rauscher13}, which is opposite to our results. Hence, this is likely not the reason for the differences. Nevertheless, our large error bars make both measurements fully compatible. In addition, the cross-correlation signal of the 1.15\,$\mu$m band is rather weak and, thus, the significance of its wind measurement is low. 
\cite{Snellen10} reported, from observations of a transit of this planet in the 2-$\mu$m spectral band of carbon monoxide, an overall blueshift of $-$2$\pm$1\,\kms. Our best measurement of the net blueshift obtained when considering all useful orders is larger, although our uncertainties also exceed theirs and, overall, both determinations are within the errors.

The model of \citet{Showman2013} predicted that the atmosphere of \hd20, which is exposed to a large stellar insolation, is expected to have a circulation dominated by high-altitude, day-to-night winds, leading to a predominantly blueshifted Doppler signature at the terminator (i.e., when probed during transit observations). The same model predicted for this planet a stronger high-altitude day-to-night circulation than for \hdu18, which should manifest in a larger net blueshift. 
In particular, it predicted a blueshifted signal with a peak near $- 3$\,\kms\ for a fraction of $\sim$40\% of the full terminator for \hdu18 at a pressure level of 0.1\,mbar, whilst for \hd20 the velocity is around $-6$\,\kms\ and covering near 60\% of the terminator (see their Figs.\,7c and 7d). These supersonic wind velocities were also obtained in the models of \citet{Rauscher12} and \citet{Amundsen16}. \citet{Rauscher13}, however, reported slightly slower winds because they included the effects of magnetic drag, which reduces the wind velocity in the upper atmosphere of \hd20 to about $\simeq$ 3 kms$^{-1}$. When considering all the useful NIR spectral orders, we obtained a blue velocity of $-$5.2\,$^{+2.6}_{-1.3}$\,\kms\ for \hd20. This value is in good agreement with the predictions of \citet{Rauscher12}, \citet{Showman2013}, and \citet{Amundsen16}. In addition, the absolute value is larger than what we derived for \hdu18 \cite[$-$3.9\,$\pm 1.3$\,kms$^{-1}$; ][]{Alonso19}, which is in line with the model predictions of \citet{Showman2013}. However, the rather large error bars in both measurements prevented us from confirming the prediction of \citet{Showman2013} of stronger winds in the upper atmosphere of \hd20.

\begin{table}
{\tiny
\centering
\caption{\label{table.global_results_hd18}Maximum S/N of the CCFs at several wavelength intervals for \hdu18.}
\begin{tabular}{lccc} 
\hline
\hline
\noalign{\smallskip}
Band & S/N & $K_{\rm P}$ [km\,s$^{-1}$] & $\varv_{\rm wind}$ [km\,s$^{-1}$] \\
\noalign{\smallskip}
\hline
\noalign{\smallskip}
1.0\,$\mu$m$^{a}$ & 3.7 (2.2) & 179\,$^{+27}_{-34}$ (153) & --3.9\,$\pm$\,2.6 ($-$3.9)\\
  \noalign{\smallskip}
1.15\,$\mu$m & 4.6 (4.9) & 116\,$^{+65}_{-31}$ (146\,$\pm$\,46) & --5.2\,$^{+2.6}_{-1.3}$ (--3.9\,$^{+1.3}_{-2.6}$)\\
 \noalign{\smallskip}
1.4\,$\mu$m & 5.1 (4.4) & 150\,$\pm$\,29 (156\,$^{+39}_{-31}$) & $-$3.9\,$\pm$\,2.6 (--3.9\,$\pm$\,1.3) \\ 
\noalign{\smallskip}
Useful orders$^{b}$  & 7.7 (6.6) & 147\,$^{+33}_{-28}$ (160\,$^{+45}_{-33}$) & --3.9\,$\pm$\,1.3 ($-$3.9\,$\pm$\,1.3)\\
\noalign{\smallskip}
\hline
\end{tabular}
\tablefoot{Values in parenthesis are from  \cite{Alonso19}. $^{a}$Value not representing the maximum of the explored map. $^{b}$Combination of the 22 useful NIR spectral orders from \cite{Alonso19}.}
}
\end{table}

\subsection{Comparison with the CARMENES \h2o detection in \hdu18}
\label{comparison_Alonso19}

We present a qualitative comparison of the results discussed above for \hd20 with those obtained by \cite{Alonso19} for the hot Jupiter \hdu18 using also CARMENES transit data (mean S/N of the continuum $\sim$\,150). In order to make it meaningful, though, we have reanalyzed the \hdu18 dataset using the same procedure used in this work. The multi-band results are shown in Fig.\,\ref{multiband_maps_hd18} and Table\,\ref{table.global_results_hd18}. 

Starting with the 1.0\,$\mu$m band, water vapor was not detected in \hdu18 from this band by \cite{Alonso19}. In this reanalysis (see left panel in Fig.\,\ref{multiband_maps_hd18}) we observe a strong contamination at low \kp\ values but also hints of a moderate signal of S/N\,$=$\,3.7 at \kp\,$=$\,179\,$^{+27}_{-34}$\,\kms, which is consistent with the expected velocities for \hdu18. Regarding the 1.15\,$\mu$m and 1.4\,$\mu$m bands, the results of the reanalysis (see Table\,\ref{table.global_results_hd18}) are very similar to those presented by \cite{Alonso19}, with confident detections of \h2o in both bands (see middle and right panels in Fig.\,\ref{multiband_maps_hd18}, respectively).
When using all the useful NIR orders from \cite{Alonso19}, we detected \h2o in \hdu18 with a S/N\,=\,7.7 at \kp\ =\,147\,$^{+33}_{-28}$\,\kms, which represents a small improvement with respect to the signal obtained in that work of S/N\,=\,6.6. This is due to the optimization of the number of {\sc Sysrem} iterations required for each spectral order, principally for the 1.0\,$\mu$m band. 

Compared to the results obtained for \hd20, the hint of a water vapor detection using the 1.0\,$\mu$m band in \hdu18 is much less significant, despite the larger S/N of the gathered spectra in the latter. This could be explained by different atmospheric extinction levels in these planets. \hdu18, compared to \hd20, is expected to be a rather hazy planet with a steeper Rayleigh scattering slope \citep{Sing16} and, hence, the extinction near 1\,$\mu$m could contribute to weaken the \h2o absorbing lines in \hdu18.

As for the 1.15\,$\mu$m and 1.4\,$\mu$m bands, \h2o in \hd20 is hinted only from the 1.15\,$\mu$m band and it is not detected from the 1.4\,$\mu$m band.
However, water vapor was detected from both bands individually and with confidence in \hdu18 by \cite{Alonso19}, as well as in this reanalysis. 
As discussed above, the lower and wavelength-dependent S/N of the measured spectra of HD\,209458 and the more variable observing conditions might be the reasons behind this difference.

\section{Conclusions} \label{conclusions}

We present the detection of water vapor in the atmosphere of the hot Jupiter \hd20 from CARMENES high-resolution NIR spectra. When including all the useful spectral orders, the \h2o signal was detected with a S/N of 6.4 and a $\sigma$-value of 8.1. We found that the largest contribution to the detection comes from the 1.0\,$\mu$m band, which is individually detected with a S/N of 5.8. This result constitutes the first detection of \h2o from this band individually. We also observed hints of a water vapor signal from the 1.15\,$\mu$m band with a low S/N of 2.8. Despite the 1.4\,$\mu$m band being the strongest one covered by the instrument, we did not find any conclusive signal from this spectral interval.

We measured a net blueshift of --5.2\,$^{+2.6}_{-1.3}$\,km\,s$^{-1}$ in the \h2o signal when using all the NIR orders expected to carry a significant signal. This indicates an average day-to-nightside strong wind at the terminator of \hd20. This wind velocity is slightly larger than that reported previously, although comparable within the uncertainty intervals. Moreover, this result supports the model predictions of \cite{Showman2013}, where a stronger day-to-nightside thermospheric wind is predicted for this hotter exoplanet than for \hdu18.

The removal of the telluric contamination is probably the most delicate step of the cross-correlation technique applied in this study. We used {\sc Sysrem}, widely tested for this purpose in previous works. We optimized the number of iterations required for each order independently by studying the behavior of an injected model signal of the planet. During the observations of the \hd20 transit, a sudden drop in the column depth of PWV occurred. Moreover, a significant portion of the observations was taken at large airmasses, resulting in low S/N spectra for the second half of the gathered spectra. In addition, the S/N of the spectra shows a strong wavelength dependence towards the end of the night, with the S/N decreasing significantly at the reddest wavelengths. These effects combined could be behind the less effective telluric removal in the two stronger bands covered, which would explain the lower than expected significances obtained from the 1.15\,$\mu$m and 1.4\,$\mu$m bands.

Our comparison of the CARMENES data for \hd20 and \hdu18 supports the previously reported hazy/cloudy nature of the latter. As discussed by \cite{Pino18b} and \cite{Alonso19}, the \h2o molecular bands studied here could ideally be used, along with optical data, to characterize the aerosol extinction in the atmospheres of hot Jupiters. However, the low CCF contrast ratios in the bands covered by the CARMENES NIR channel and the atmospheric variability in the \hd20 observation night made such a study unfeasible in this case. Future observations under more favorable conditions might allow us to detect \h2o from all the bands covered by the CARMENES NIR channel and also from the weaker bands in the VIS channel, and hence to tackle such study.

\begin{acknowledgements} 
CARMENES is funded by the German Max-Planck-Gesellschaft (MPG), the Spanish Consejo Superior de Investigaciones Cient\'ificas (CSIC), the European Union through European Regional Fund (FEDER/ERF), the Spanish Ministry of Economy and Competitiveness, the state of Baden-W\"urttemberg, the German Science Foundation (DFG), and the Junta de Andaluc\'ia, with additional contributions by the members of the CARMENES Consortium (Max-Planck-Institut f\"ur Astronomie, Instituto de Astrof\'isica de Andaluc\'ia, Landessternwarte K\"onigstuhl, Institut de Ci\`encies de l'Espai, Institut f\"ur Astrophysik G\"ottingen, Universidad Complutense de Madrid, Th\"uringer Landessternwarte Tautenburg, Instituto de Astrof\'isica de Canarias, Hamburger Sternwarte, Centro de Astrobiolog\'ia, and the Observatorio de Calar Alto).
Financial support was also provided by the {Universidad Complutense de Madrid}, the Comunidad Aut\'onoma de Madrid, the Spanish Ministerios de Ciencia e Innovaci\'on and of Econom\'ia y Competitividad, the Fondo Europeo de Desarrollo Regional (FEDER/ERF), the Agencia estatal de investigaci{\'o}n, and the Fondo Social Europeo under grants 
AYA2011-30147-C03-01, -02, and -03, 
AYA2012-39612-C03-01, 
ESP2013-48391-C4-1-R, 
ESP2014--54062--R,
ESP 2016--76076--R, ESP2016-80435-C2-2-R, ESP2017-87143-R, BES--2015--073500,
and BES--2015--074542. 
IAA authors acknowledge financial support from the State Agency for Research of the Spanish MCIU through the ``Center of Excellence Severo Ochoa" award SEV-2017-0709. L.T.-O. acknowledges support from the Israel Science Foundation (grant No. 848/16).  
Based on observations collected at the Observatorio de Calar Alto. 
We thank the anonymous referee for their insightful comments, which contributed to improve the quality of the manuscript.
\end{acknowledgements}


\end{document}